\documentclass[aps,prd,twocolumn,superscriptaddress,amsmath,amssymb]{revtex4}
\usepackage{graphicx} 
\usepackage{epstopdf} 
\usepackage{dcolumn} 
\usepackage{xcolor} 
\usepackage{amsmath,amssymb,amsfonts,mathrsfs} 
\usepackage{hyperref} 
\usepackage[utf8]{inputenc} 
\usepackage[english]{babel} 
\usepackage[mathlines]{lineno} 
\usepackage{blindtext} 
\usepackage{ifthen} 
\usepackage{orcidlink} 
\usepackage{tabularx}

\usepackage{booktabs}
\AtBeginDocument{%
  \heavyrulewidth=.08em
  \lightrulewidth=.05em
  \cmidrulewidth=.03em
  \belowrulesep=.65ex
  \belowbottomsep=0pt
  \aboverulesep=.4ex
  \abovetopsep=0pt
  \cmidrulesep=\doublerulesep
  \cmidrulekern=.5em
  \defaultaddspace=.5em
}

\usepackage[noabbrev,capitalise]{cleveref} 

\usepackage{afterpage} 
\usepackage{rotating,booktabs,multirow} 

\usepackage[percent]{overpic}

\graphicspath{{figures/}} 

\newboolean{articletitles}
\setboolean{articletitles}{true} 

\newboolean{twocolumnstyle}
\setboolean{twocolumnstyle}{false}

\newcommand*\patchAmsMathEnvironmentForLineno[1]{%
\expandafter\let\csname old#1\expandafter\endcsname\csname #1\endcsname
\expandafter\let\csname oldend#1\expandafter\endcsname\csname
end#1\endcsname
 \renewenvironment{#1}%
   {\linenomath\csname old#1\endcsname}%
   {\csname oldend#1\endcsname\endlinenomath}%
}
\newcommand*\patchBothAmsMathEnvironmentsForLineno[1]{%
  \patchAmsMathEnvironmentForLineno{#1}%
  \patchAmsMathEnvironmentForLineno{#1*}%
}
\AtBeginDocument{%
\patchBothAmsMathEnvironmentsForLineno{equation}%
\patchBothAmsMathEnvironmentsForLineno{align}%
\patchBothAmsMathEnvironmentsForLineno{flalign}%
\patchBothAmsMathEnvironmentsForLineno{alignat}%
\patchBothAmsMathEnvironmentsForLineno{gather}%
\patchBothAmsMathEnvironmentsForLineno{multline}%
\patchBothAmsMathEnvironmentsForLineno{eqnarray}%
}

\input{belle2-symbols}

\setboolean{twocolumnstyle}{true}

\newcommand{\sub}[1]{\ensuremath{_{\text{#1}}}\xspace}
\newcommand{\range}[3]{\ensuremath{[#1, #2]#3}\xspace}

\newcommand{\Acp}{\ensuremath{A_{\CP}}\xspace}
\newcommand{\Araw}{\ensuremath{A_{N}}\xspace}

\newcommand{\Xc}{\ensuremath{\Xires^{+}_c}\xspace}

\newcommand{\Sigp}{\ensuremath{\Sigmares^{+}}\xspace}
\newcommand{\p}{\ensuremath{p}\xspace}

\newcommand{\hphm}{\ensuremath{h^{+}h^{-}}\xspace}
\newcommand{\Shh}{\ensuremath{\Sigp\hphm}}
\newcommand{\phh}{\ensuremath{\p\hphm}}
\newcommand{\SKK}{\ensuremath{\Sigp\Kp\Km}\xspace}
\newcommand{\Spipi}{\ensuremath{\Sigp\pip\pim}\xspace}
\newcommand{\pKK}{\ensuremath{\p\Kp\Km}\xspace}
\newcommand{\ppipi}{\ensuremath{\p\pip\pim}\xspace}

\newcommand{\XcToShh}{\ensuremath{\Xc\to\Shh}}
\newcommand{\LcTophh}{\ensuremath{\Lc\to\phh}}
\newcommand{\XcToSKK}{\ensuremath{\Xc\to\SKK}}
\newcommand{\XcToSpipi}{\ensuremath{\Xc\to\Spipi}}
\newcommand{\LcToShh}{\ensuremath{\Lc\to\Shh}}
\newcommand{\LcToSKK}{\ensuremath{\Lc\to\SKK}}
\newcommand{\LcToSpipi}{\ensuremath{\Lc\to\Spipi}}
\newcommand{\LcTopKK}{\ensuremath{\Lc\to\pKK}}
\newcommand{\LcToppipi}{\ensuremath{\Lc\to\ppipi}}
\newcommand{\LcTopKpi}{\ensuremath{\Lc\to\p\pip\Km}\xspace}
\newcommand{\DToKpi}{\ensuremath{\Dz\to\pip\Km}\xspace}
\newcommand{\DToKpipipi}{\ensuremath{\Dz\to\pip\Km\pipi}\xspace}

\newcommand{\lumi}{\ensuremath{428\invfb}\xspace}

\newcommand{\AcpSKK}{3.7}
\newcommand{\AcpSKKStat}{6.6}
\newcommand{\AcpSKKSyst}{0.6}

\newcommand{\AcpSpipi}{9.5}
\newcommand{\AcpSpipiStat}{6.8}
\newcommand{\AcpSpipiSyst}{0.5}

\newcommand{\AcppKK}{3.9}
\newcommand{\AcppKKStat}{1.7}
\newcommand{\AcppKKSyst}{0.7}

\newcommand{\Acpppipi}{0.3}
\newcommand{\AcpppipiStat}{1.0}
\newcommand{\AcpppipiSyst}{0.2}

\newcommand{\kekpreprint}{2025-26}  
\newcommand{\biipreprint}{2025-024}  

\newcommand{\asym}[2]{\ensuremath{\frac{#1 - #2}{#1 + #2}}\xspace}

\begin{document}

\includegraphics[width=3cm]{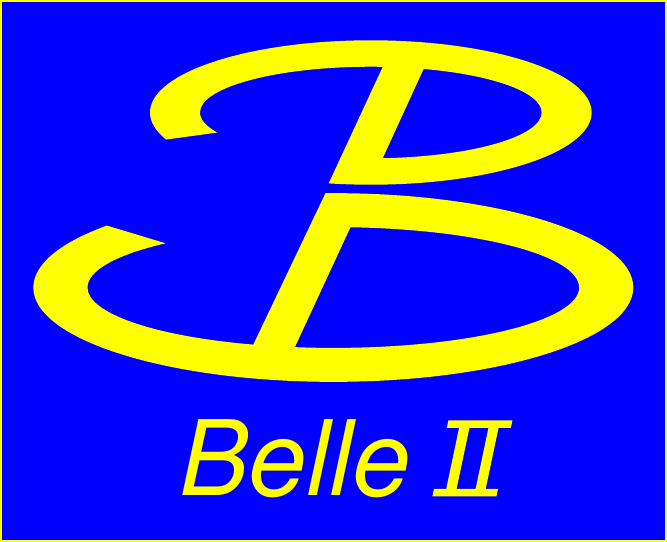}\vspace*{-1.9cm}
\begin{flushright}

Belle II Preprint \biipreprint

KEK Preprint \kekpreprint
\end{flushright}\vspace{1.5cm}

\title{{\LARGE\bfseries\boldmath Search for \CP violation in \XcToShh\ and \LcTophh\ at Belle~II}}
  \author{M.~Abumusabh\,\orcidlink{0009-0004-1031-5425}} 
  \author{I.~Adachi\,\orcidlink{0000-0003-2287-0173}} 
  \author{H.~Ahmed\,\orcidlink{0000-0003-3976-7498}} 
  \author{Y.~Ahn\,\orcidlink{0000-0001-6820-0576}} 
  \author{H.~Aihara\,\orcidlink{0000-0002-1907-5964}} 
  \author{N.~Akopov\,\orcidlink{0000-0002-4425-2096}} 
  \author{S.~Alghamdi\,\orcidlink{0000-0001-7609-112X}} 
  \author{M.~Alhakami\,\orcidlink{0000-0002-2234-8628}} 
  \author{N.~Althubiti\,\orcidlink{0000-0003-1513-0409}} 
  \author{K.~Amos\,\orcidlink{0000-0003-1757-5620}} 
% ...
\collaboration{The Belle II Collaboration}

\begin{abstract}
We report decay-rate \CP asymmetries of the singly-Cabibbo-suppressed decays \XcToShh and \LcTophh, with $h=K,\pi$, measured using \lumi of \epem collisions collected by the \belletwo experiment at the SuperKEKB collider. The results,
\begin{alignat*}{1}
    \Acp(\XcToSKK)   &= (\AcpSKK\pm\AcpSKKStat\pm\AcpSKKSyst)\%, \\ 
    \Acp(\XcToSpipi) &= (\AcpSpipi\pm\AcpSpipiStat\pm\AcpSpipiSyst)\%, \\
    \Acp(\LcTopKK)   &= (\AcppKK\pm\AcppKKStat\pm\AcppKKSyst)\%, \\
    \Acp(\LcToppipi) &= (\Acpppipi\pm\AcpppipiStat\pm\AcpppipiSyst)\%,
\end{alignat*}
where the first uncertainties are statistical and the second systematic, agree with \CP symmetry. From these results we derive the sums
\begin{alignat}{3}
  &\Acp(\XcToSpipi) &&\, + \, \Acp(\LcTopKK)   &&= (13.4 \pm 7.0\pm 0.9)\%,\\
  &\Acp(\XcToSKK)   &&\, + \, \Acp(\LcToppipi) &&= (\phantom{0}4.0 \pm 6.6\pm 0.7)\%,
\end{alignat}
which are consistent with the $U$-spin symmetry prediction of zero. These are the first measurements of \CP asymmetries for individual hadronic three-body charmed-baryon decays.
\end{abstract}

\maketitle

\section{Introduction}
Violation of charge-parity (\CP) symmetry has been established in decays of strange, charmed, and bottom mesons~\cite{C64,KTeV:1999kad,NA48:2001bct,BaBar:2001pki,Belle:2001zzw,BaBar:2004gyj,Belle:2004nch,LHCb:2013syl,LHCb:2012xkq,A19,A23} and recently bottom baryons~\cite{A25}. The LHCb collaboration established \CP violation in charm decays by observing a nonzero difference between the decay-rate \CP asymmetries of $\Dz\to\Kp\Km$ and $\Dz\to\pip\pim$~\cite{A19}. (Charge-conjugate modes are included implicitly.) The standard model explains the measured \CP asymmetries of strange and bottom decays  (see, for example, Sections 12 and 13 of Ref.~\cite{pdg}), yet offers no clear interpretation of the \CP asymmetry measured for charm decays. Some theoretical interpretations have suggested the measured \CP asymmetry of order $10^{-3}$ arises from interactions beyond the standard model~\cite{Chala:2019,Dery:2019ysp,Calibbi:2019bay,Buras:2021rdg,Pich:2023kim,Lenz:2023rlq,Sinha:2025}, others from nonperturbative QCD~\cite{Grossman:2019,Cheng:2019ggx,Schacht:2021jaz,Bediaga:2022sxw,Gavrilova:2023fzy}. Moreover, LHCb's separate determinations of $\Acp(\Dz\to\Kp\Km)$ and $\Acp(\Dz\to\pip\pim)$~\cite{A23} indicate unexpectedly large breaking of $U$-spin symmetry~\cite{Schacht:2022kuj}, which assumes invariance under the exchange of down and strange quarks and predicts that the \CP asymmetries of $D^0\to\Kp\Km$ of and $D^0\to\pip\pim$ have equal magnitudes and opposite signs~\cite{Buccella:1994nf,Grossman:2006jg}.

Another place to search for \CP violation and test $U$-spin symmetry is in three-body, singly-Cabibbo-suppressed, hadronic charmed-baryon decays~\cite{Grossman:2018ptn,Wang:2019dls}. Assuming $U$-spin symmetry,
\begin{alignat}{3}
    &\Acp(\XcToSpipi) &&\ +\ \Acp(\LcTopKK) &&= 0, \label{uspinrule1}\\
    &\Acp(\XcToSKK)   &&\ +\ \Acp(\LcToppipi) &&= 0. \label{uspinrule2}
\end{alignat}
None of these asymmetries have been measured, individually or in these sums, despite the large branching fractions of their decays. LHCb measured the difference between $\Acp(\LcTopKK)$ and $\Acp(\LcToppipi)$~\cite{aaij_measurement_2018}, but since they are not related by $U$-spin this does not reveal anything about $U$-spin-symmetry breaking.

We report the first measurements of the \CP asymmetries of \XcToShh and \LcTophh, with $h=K,\pi$. We use \lumi of \epem collisions collected at or near the $\Upsilon(4S)$ center-of-mass energy by the \belletwo experiment from 2019 to 2022.

The decay-rate \CP asymmetry of charmed baryon $X_c^+$ decaying to final state $f^+$ is
\begin{equation}
    \Acp(X_c^+ \to f^+) \equiv \asym{\Gamma(X_c^+ \to f^+)}{\Gamma(\bar{X}_c^- \to \bar{f}^-)}.
\end{equation}
We determine it by measuring the asymmetry of yields,
\begin{equation}
    \Araw(X_c^+ \to f^+) \equiv \asym{N(X_c^+ \to f^+)}{N(\bar{X}_c^- \to \bar{f}^-)},
    \label{eq:araw}
\end{equation}
which we correct for detection and production asymmetries. In the limit of small asymmetries, 
\begin{equation}
    \Araw(X_c^+ \to f^+) = \Acp(X_c^+ \to f^+) \, + \, A\sub{p}(X_c^+) \, + \, A\sub{d}(f^+),
\end{equation}
where $A\sub{p}(X_c^+)$ is due to the forward-backward asymmetry of charmed-baryon production in \epem collisions~\cite{Berends:1973fd,Brown:1973ji,Cashmore:1985vp} and $A\sub{d}(f^+)$ is the asymmetry of detecting the charged final-state particles. 

Since $A\sub{p}$ is an odd function of the cosine of the polar angle of the charmed baryon's momentum in the \epem center-of-mass~(c.m.)\ frame, $\cos\theta$, we suppress it by replacing \Araw in \cref{eq:araw} with the average of the \Araw measured separately in the forward and backward regions,
\begin{equation}
    \Araw^{\prime}\!=\! \frac{\Araw(\cos\theta>0) + \Araw(\cos\theta<0)}{2}.
\end{equation} 
We remove the charged-particle detection asymmetries by subtracting the yield asymmetries of control channels, \LcToShh, \LcTopKpi, and \DToKpipipi. We use \DToKpipipi rather than \DToKpi because the phase-space of the \Km\pip pair is similar for \LcTopKpi and \DToKpipipi, but not for \DToKpi. Since these decays are Cabibbo favored, they have high yields and are expected to have negligible \CP asymmetries, which we assume are zero. The decay-rate \CP asymmetry of \XcToShh is 
\ifthenelse{\boolean{twocolumnstyle}}{
\begin{align}
\Acp(\XcToShh) &= \Araw'(\XcToShh) \nonumber\\ &- \Araw'(\LcToShh),
\label{AcpXic}
\end{align}
}{
\begin{equation}
\Acp(\XcToShh) = \Araw'(\XcToShh) - \Araw'(\LcToShh),
\label{AcpXic}
\end{equation}
}
and that of \LcTophh is
\ifthenelse{\boolean{twocolumnstyle}}{
\begin{align}
\Acp(\LcTophh) &= \Araw'(\LcTophh) \nonumber\\ &- \Araw'(\LcTopKpi) \nonumber\\ &+ \Araw'(\DToKpipipi).
\label{AcpLc}
\end{align}
}{
\begin{equation}
\Acp(\LcTophh) = \Araw'(\LcTophh) -\Araw'(\LcTopKpi) + \Araw'(\DToKpipipi).
\label{AcpLc}
\end{equation}
}
Subtracting the \Lc asymmetries cancels final-state-baryon detection asymmetries. Adding the \Dz asymmetry cancels \Km and \pip detection asymmetries that the subtraction of $\Araw'(\LcTopKpi)$ introduces. For the former, the distributions of the final-state baryon's c.m.~frame momentum and $\cos\theta$ must be the same in the signal and control channels. For the latter, the same must be true of the kaon-pion pair. We apply kinematic weights to candidates in each control channel so that their distributions match those of their signal channel. We assume there is no asymmetry in detecting the \hphm pair from the \Xc and \Lc decays, and assign a systematic uncertainty for this assumption. The measured values of $\Araw'(\XcToShh)$ and $\Araw'(\LcTophh)$ remained unexamined until the entire analysis procedure was finalized to avoid potential bias.

\section{\belletwo detector and simulation}
\label{sec:detector}

The \belletwo detector~\cite{Abe:2010gxa}, which is located at the SuperKEKB asymmetric-energy \epem collider~\cite{Akai:2018mbz}, has a cylindrical geometry and includes a tracking system comprising a two-layer silicon pixel detector~(PXD) surrounded by a four-layer double-sided silicon strip detector~(SVD) and a 56-layer central drift chamber~(CDC), which also measures charged-particle energy loss. For the data used here, the second layer of the PXD covered only 15\% of the azimuthal range. The symmetry axis of these detectors, the $z$ axis, is nearly coincident with the direction of the electron beam. Outside the CDC is a time-of-propagation detector~(TOP) in the barrel region and an aerogel-based ring-imaging Cherenkov detector~(ARICH) in the forward endcap. They provide information for charged-particle identification together with the energy loss measurements from the SVD and CDC. Surrounding the TOP and ARICH is an electromagnetic calorimeter~(ECL) based on CsI(Tl) crystals that measures the energy deposited by photons, charged particles, \KL, and neutrons. Outside the ECL is a superconducting solenoid magnet that provides a 1.5\,T field parallel to the $z$ axis. Its flux return is instrumented with resistive-plate chambers and plastic scintillator modules to detect muons, \KL, and neutrons.

We use simulated events to identify sources of background, optimize candidate selection, weight control modes, determine fit models, and validate the analysis procedure. We generate $\epem\to\qqbar$, where $q=u,d,s,c$, with \textsc{KKMC}~\cite{Jadach:1999vf}, hadronize quarks with \textsc{Pythia}~8~\cite{Sjostrand:2014zea}, generate $\Upsilon(4S)\to B\bar{B}$ events and particle decay with \textsc{EvtGen}~\cite{Lange:2001uf} and \textsc{Pythia}~8, and simulate detector response with \textsc{Geant4}~\cite{Agostinelli:2002hh}. Beam-induced backgrounds are accounted for by including randomly triggered data. Both simulated and real data are reconstructed with the \belletwo analysis software framework~\cite{Kuhr:2018lps,basf2-zenodo}.

\section{Candidate selection}
\label{sec:selection}

Each charged particle must be associated with at least one hit in the CDC and have a distance of closest approach to the \epem interaction point~(IP) of less than 3\cm in $z$ and 1\cm in the perpendicular plane. We identify charged particles as pions, kaons, or protons using information from all subdetector systems except the PXD~\cite{PID}. A proton is identified using the likelihood of its track to be from a proton divided by the sum of likelihoods for it to be from a proton, kaon, or pion. Pions and Kaons are identified using neural-network trained particle-identification probabilities. \Sigp candidates are reconstructed via $\Sigp\to p\piz$ and \piz via $\piz\to\gamma\gamma$.  

A kinematic-vertex fit constrains the final-state particles' origin points and four-momenta to be consistent with the decay topology, the masses of the \Sigp and $\pi^0$ candidates to be their known values~\cite{pdg}, and the initial-state baryon to originate from the IP~\cite{treefitter}. Only candidates with fit $\chi^2$ probabilities greater than 0.001 are retained for subsequent analysis. Backgrounds due to random combinations of final-state particles and due to charmed baryons originating from decays of $B$ mesons are reduced by restricting the the initial-state baryon's c.m.-frame momentum and flight significance---the displacement vector from the IP to the decay vertex position, projected onto the momentum direction, divided by its uncertainty.

We optimize selection criteria by maximizing $S/\sqrt{S+B}$, where $S$ and $B$ are the signal and background yields determined from the fit described in \Cref{sec:results} in a signal channel in data in the region around the known initial-state baryon mass~\cite{pdg}. For each control channel, we use the same criteria as for its corresponding signal channel.

\subsection{\XcToShh and \LcToShh}

We reconstruct each \XcToShh or \LcToShh candidate from two oppositely charged particles and one \Sigp candidate. Protons are identified with 92\% efficiency and 5\% kaon-as-proton and 0.3\% pion-as-proton misidentification rates. Kaons and pions are identified with 89\% and 98\% efficiencies, respectively, with corresponding pion-as-kaon and kaon-as-pion misidentification rates of 7\% and 24\%.

Photon candidates are formed from clusters of energy deposition in the ECL that are not associated with a reconstructed charged particle. Each cluster must be in a region covered also by the CDC and have energy greater than 80\mev if in the ECL's forward endcap, 30\mev if in its barrel, and 60\mev if in its backward endcap. Background photons are rejected using two multivariate classifiers that determine if a cluster is due to beam-induced background or a hadronic shower~\cite{Cheema:2024iek}. They use shower-related variables, the time difference between the \epem collision and the cluster signal, and other information related to photon detection. A photon pair with mass in \range{120}{145}{\mevcc} is accepted as a neutral pion candidate.

A $p\piz$ pair with mass in \range{1.166}{1.211}{\gevcc} is accepted as a \Sigp candidate; the mass resolution is 7\mevcc. For $\Sigp\Kp\Km$, the \Sigp flight significance must be above 8 and the initial-state baryon must have c.m.-frame momentum above 2.4\gevc and flight significance above $-1.5$. For $\Sigp\pip\pim$, the \Sigp flight significance must be above 16 and the initial-state baryon must have c.m.-frame momentum above 2.7\gevc and flight significance above 3.6.

\subsection{\LcTophh and \LcTopKpi}

We reconstruct each \LcTophh or \LcTopKpi candidate from one proton candidate and two oppositely charged particles. Protons are identified with 90\% efficiency, with 2\% kaon-as-proton and 0.1\% pion-as-proton misidentification rates. Kaons and pions are identified with 79\% and 90\% efficiencies, respectively, with corresponding pion-as-kaon and kaon-as-pion misidentification rates of 2\% and 7\%. 

For $p\Kp\Km$, the \Lc must have c.m.-frame momentum above 2.3\gevc and flight significance above $-0.75$. For $p\pip\pim$, the \Lc must have c.m.-frame momentum above 2.5\gevc and flight significance above 2.0. For $p\Km\pip$, the \Lc candidate must satisfy the same momentum and flight-significance requirements as the signal channel its asymmetry is being subtracted from.

\subsection{\DToKpipipi}

We reconstruct each \DToKpipipi candidate from four charged particles with zero net charge. Charged pion and kaon candidates must satisfy the same pion and kaon criteria as for \LcTophh. The \Dz candidate must have c.m.-frame momentum above 2.3\gevc, mass in \range{1.8}{1.92}{\gevcc}, and $\chi^2$ probability above 0.001 for a kinematic-vertex fit that constrains its momentum to point back to the IP. Due to the very large yield of the \DToKpipipi control mode, we randomly select 10\% of \Dz candidates to use for the asymmetry determination.

\section{Yield and asymmetry determination}
\label{sec:results}

We determine the asymmetries from the yields of the signal and control channels, separated into the forward and backward regions and by charge, by maximizing extended likelihoods of the unbinned mass distributions of the selected decay candidates. Simultaneous fits are performed for each pair of signal and corresponding charmed-baryon control channels, separated by charge, and for the forward and backward regions. The likelihoods parameterize two sources of candidates, correctly reconstructed ones and background; the number of misreconstructed candidates is negligible according to simulation.

We model each channel's correct-candidate distribution with a double-sided Crystal Ball function~\cite{Gaiser:Phd,Skwarnicki:1986xj}, with its parameters fixed to values determined from fits to simulated data integrated over the forward and backward regions. We model each channel's background distribution with a straight line with its parameters free. The yields of correctly reconstructed candidates and background are free.

For the correct-candidate models of each pair of signal channel and charmed-baryon control channel, we add two free parameters to account for a potential common offset and a potential common scaling of widths from the values determined from fits to simulated data. They are predominantly determined from the high-yield control modes. We add a mode offset and width scaling for the \DToKpipipi channels as well, determined independently from the charmed-baryon channels.

\begin{figure*}[ht]
\centering
\includegraphics[width=0.45\linewidth]{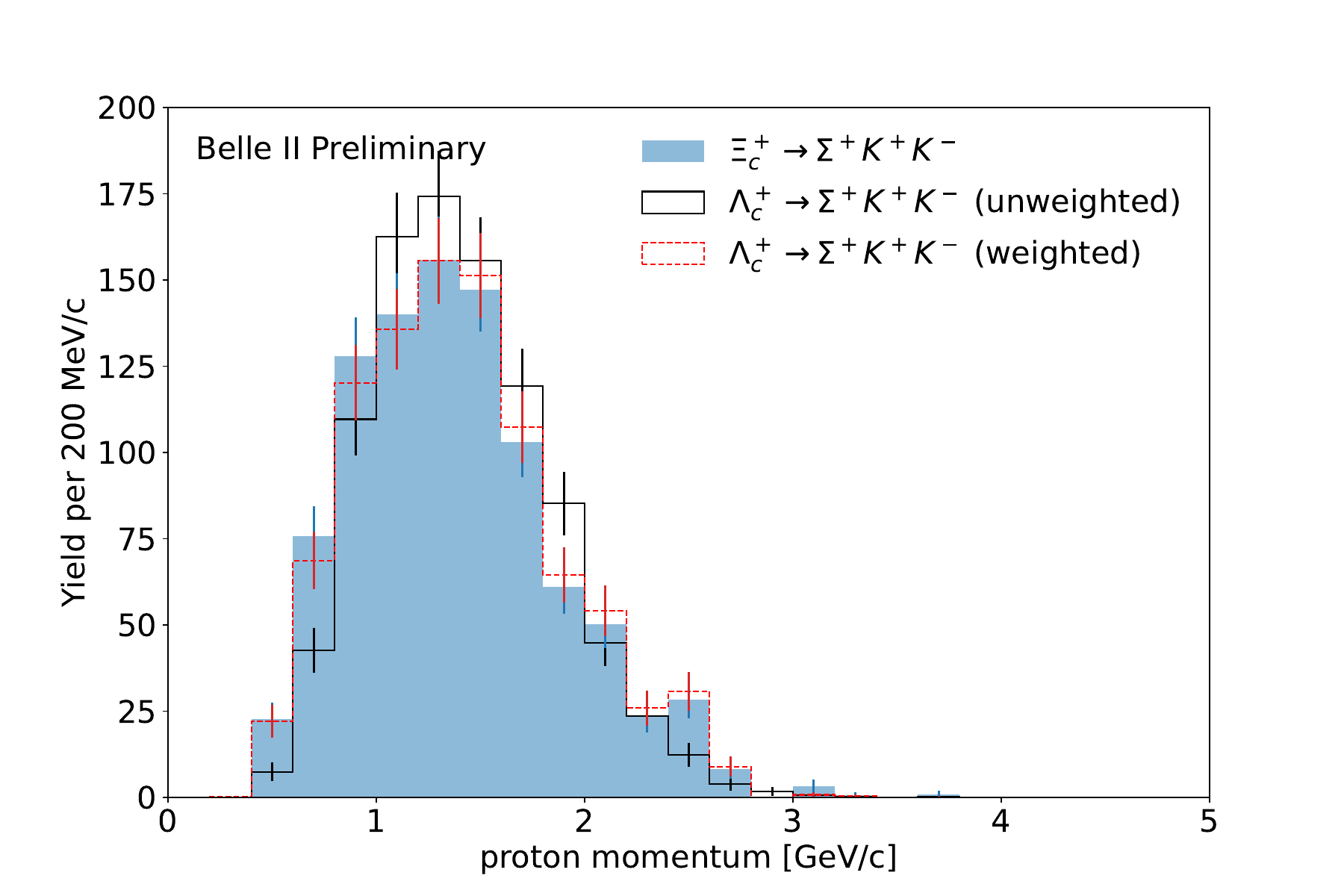}\hfil
\includegraphics[width=0.45\linewidth]{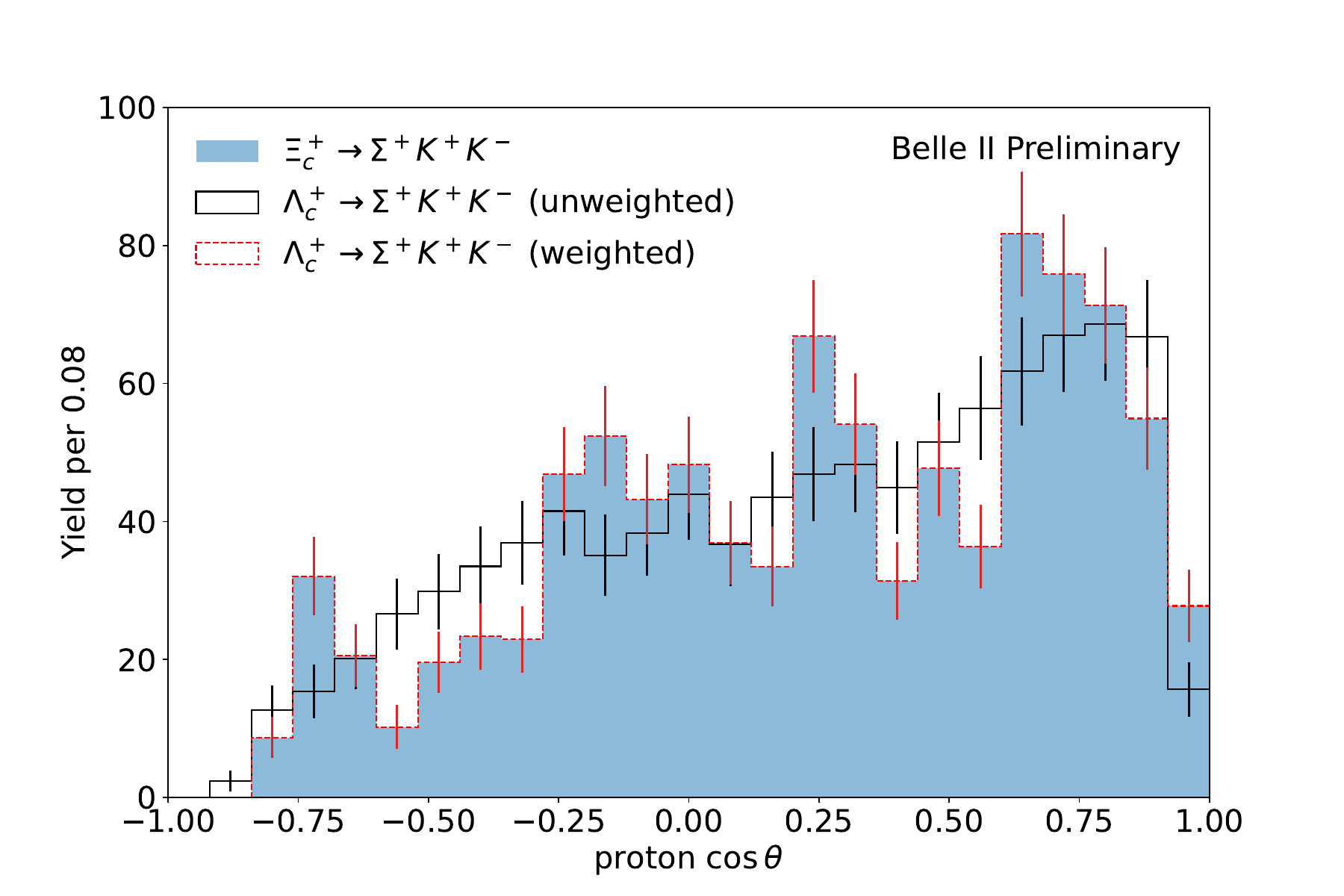}\\
\includegraphics[width=0.45\linewidth]{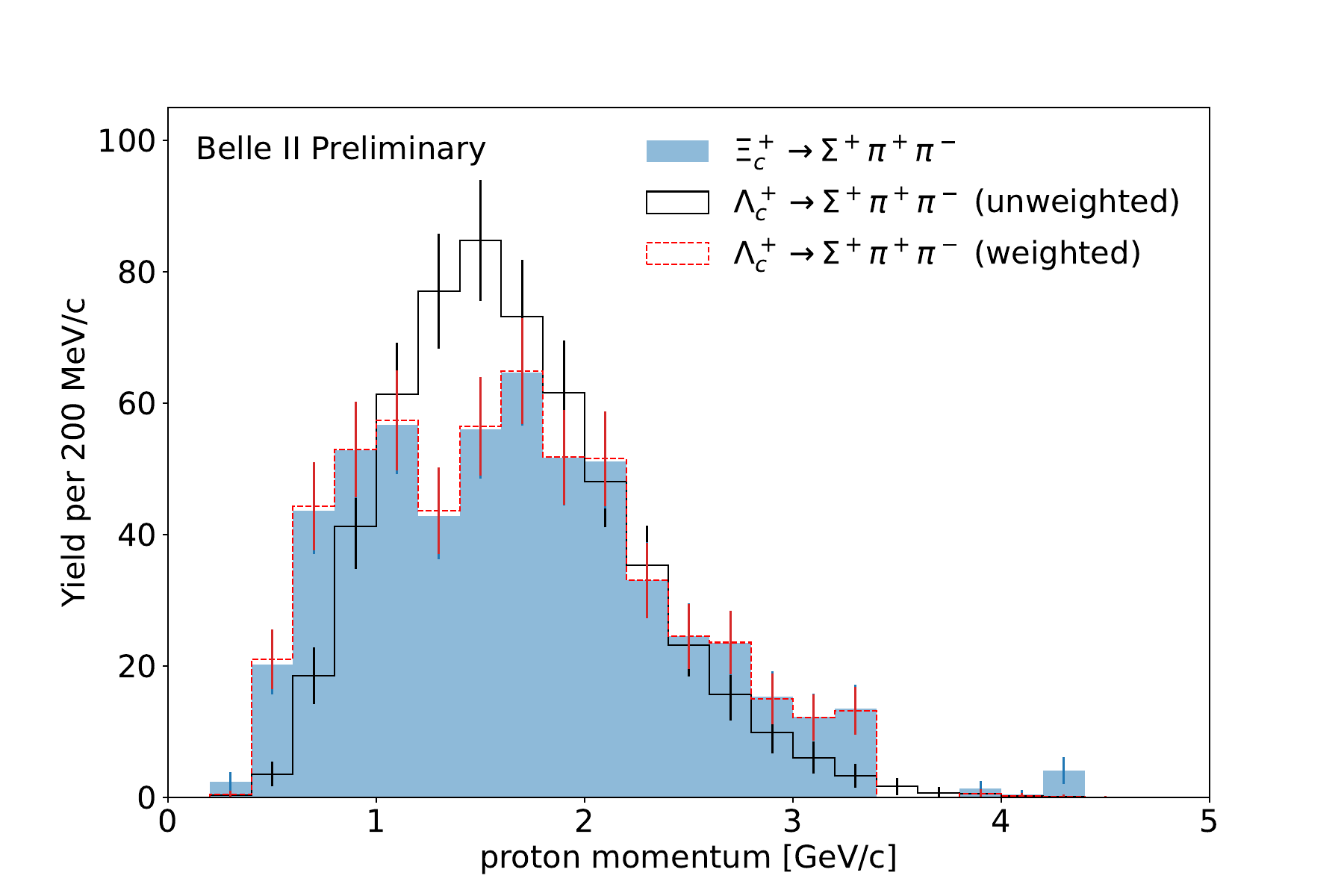}\hfil
\includegraphics[width=0.45\linewidth]{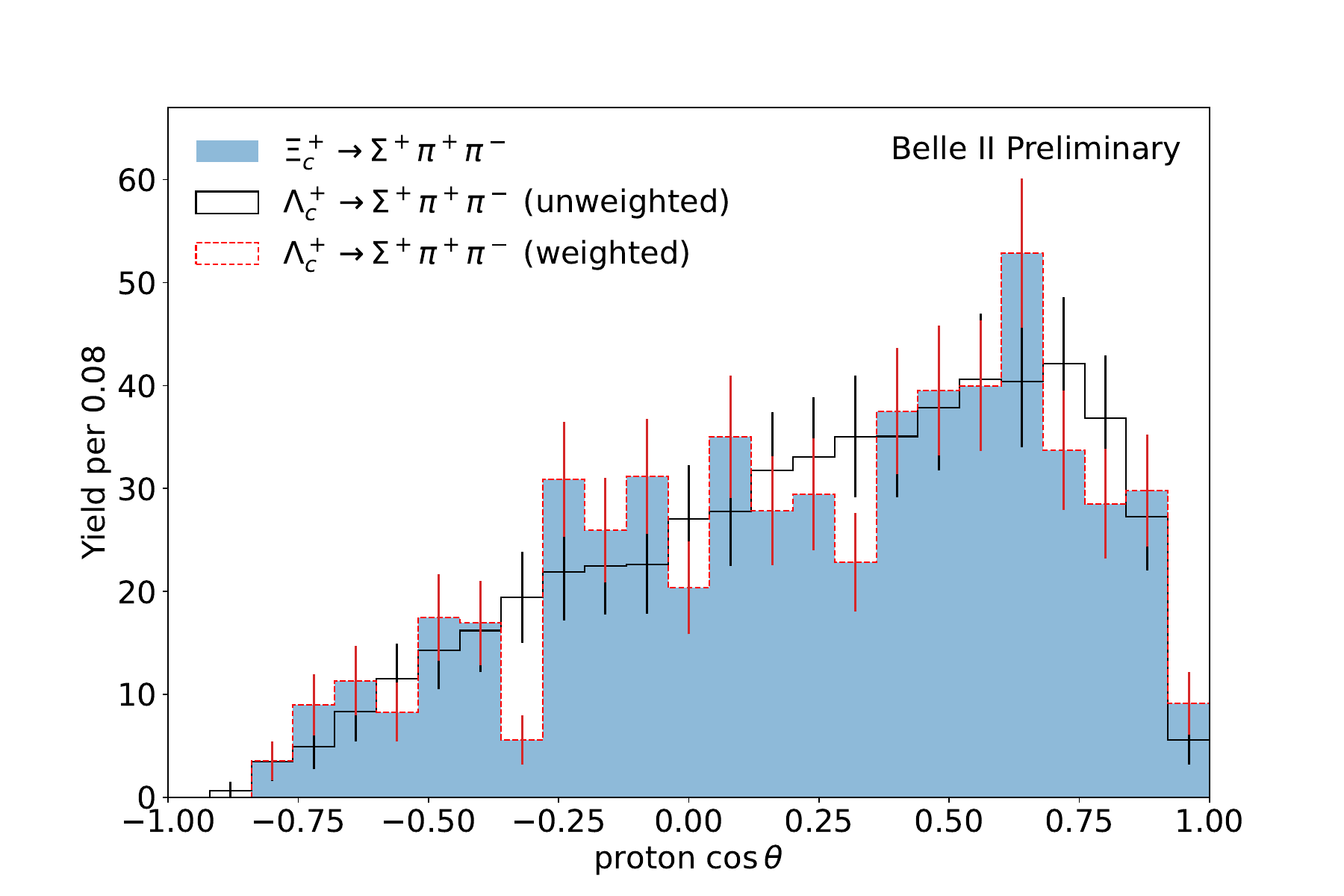}\\
\caption{Weighted and unweighted correct-candidate proton momentum~(left) and $\cos\theta$~(right) distributions for \XcToSKK and \LcToSKK~(top) and \XcToSpipi and \LcToSpipi~(bottom).}
\label{fig:ScTohh_weights}
\end{figure*}

\begin{figure*}[ht]
\centering
\includegraphics[width=0.45\linewidth]{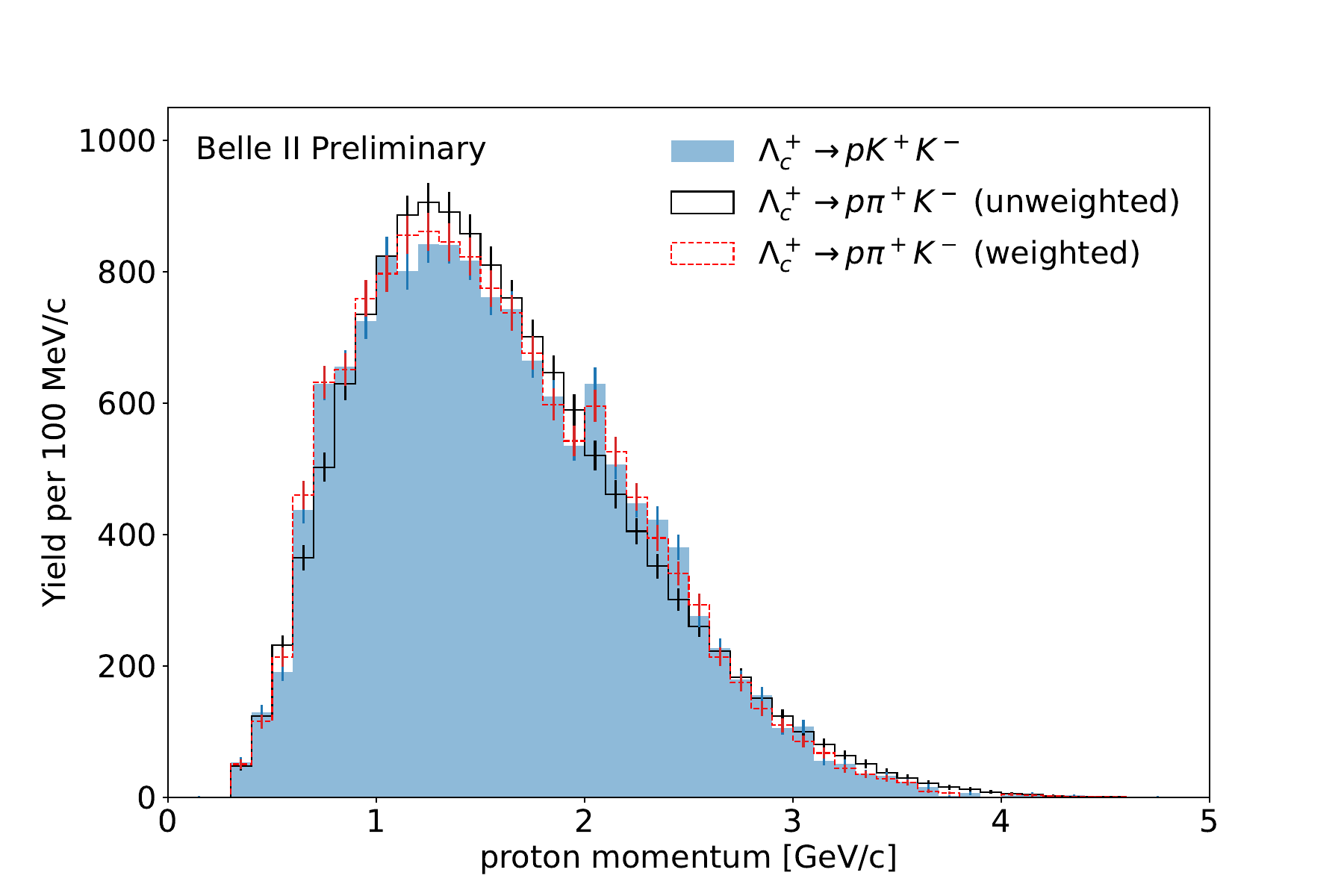}\hfil
\includegraphics[width=0.45\linewidth]{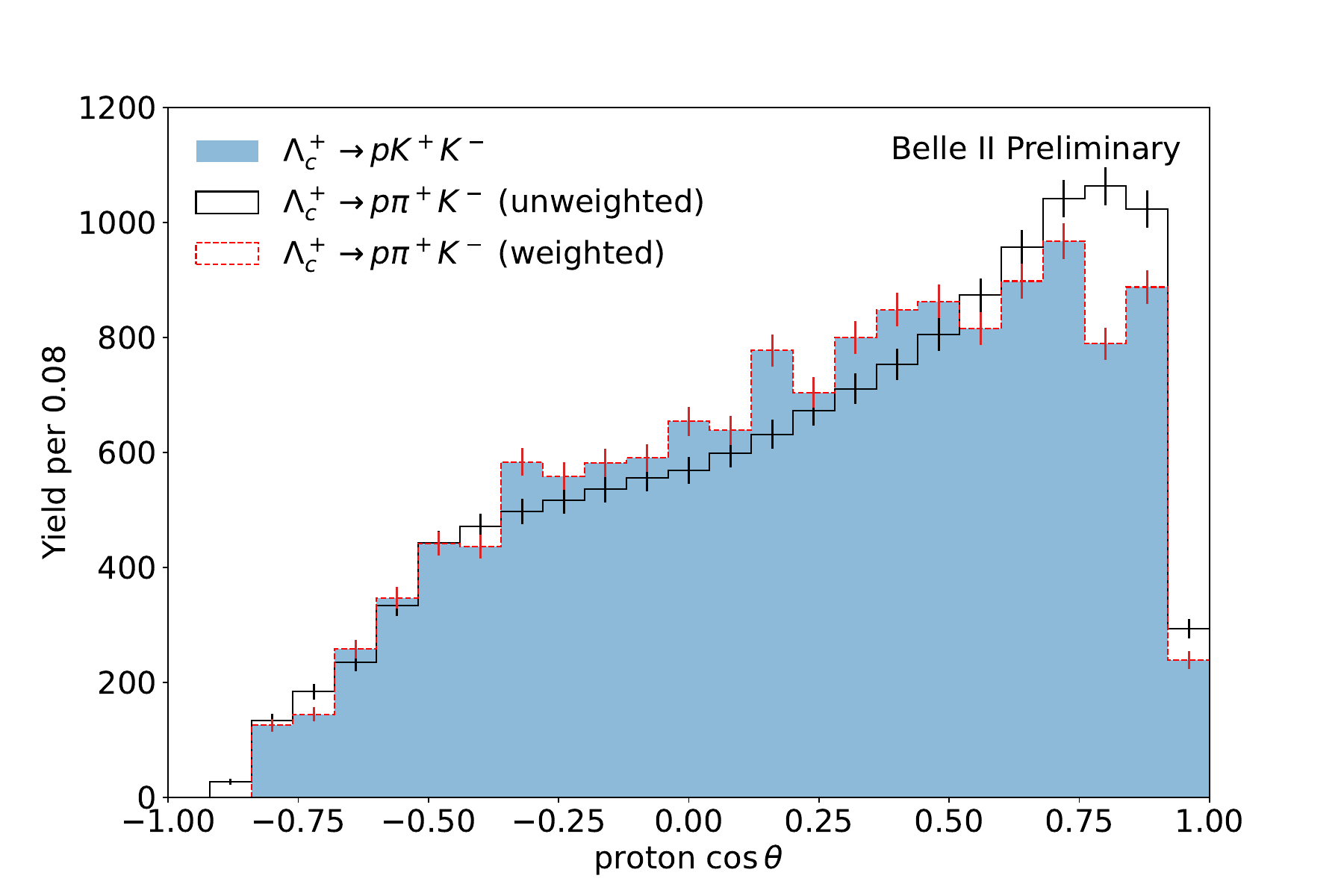}\\
\includegraphics[width=0.45\linewidth]{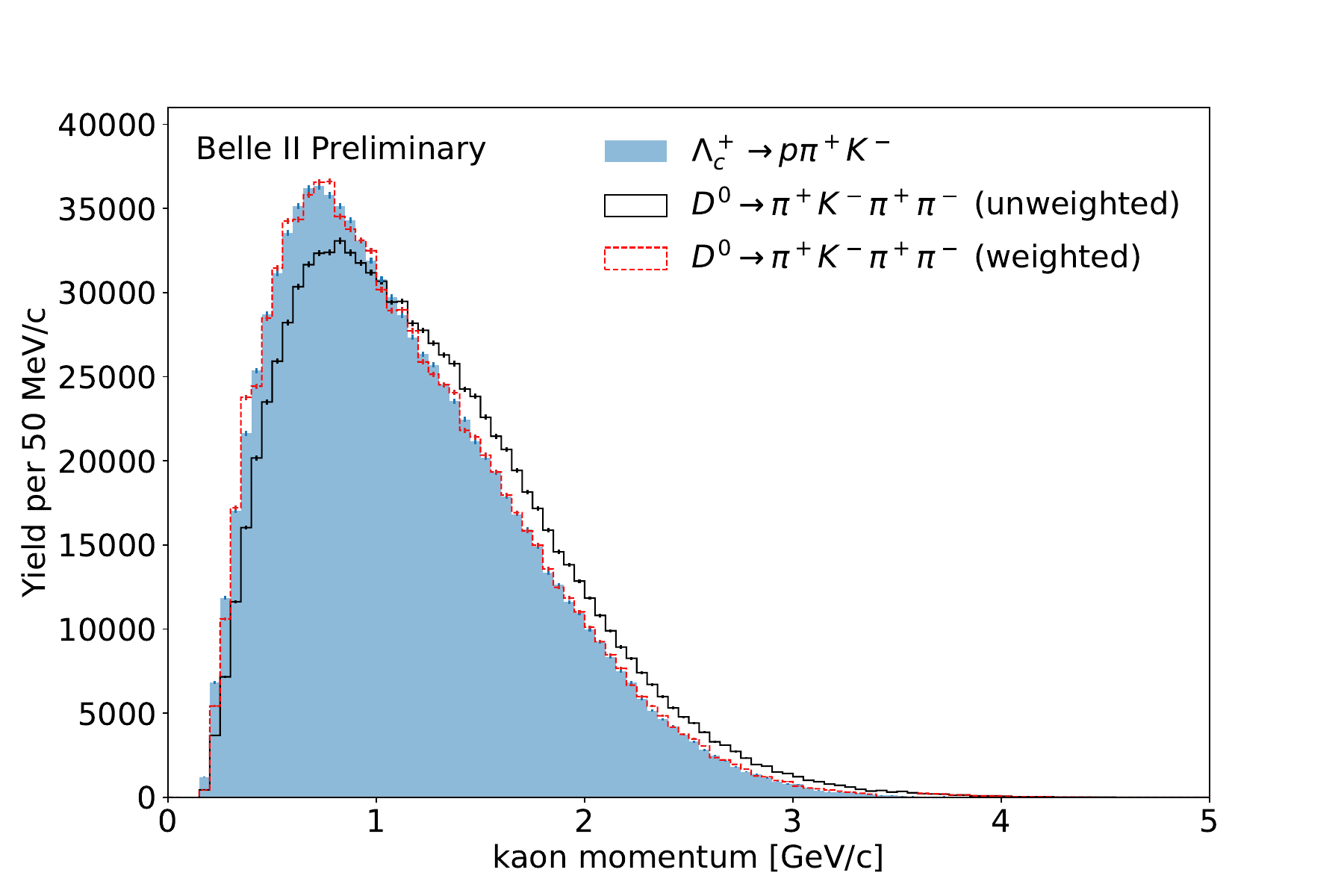}\hfil
\includegraphics[width=0.45\linewidth]{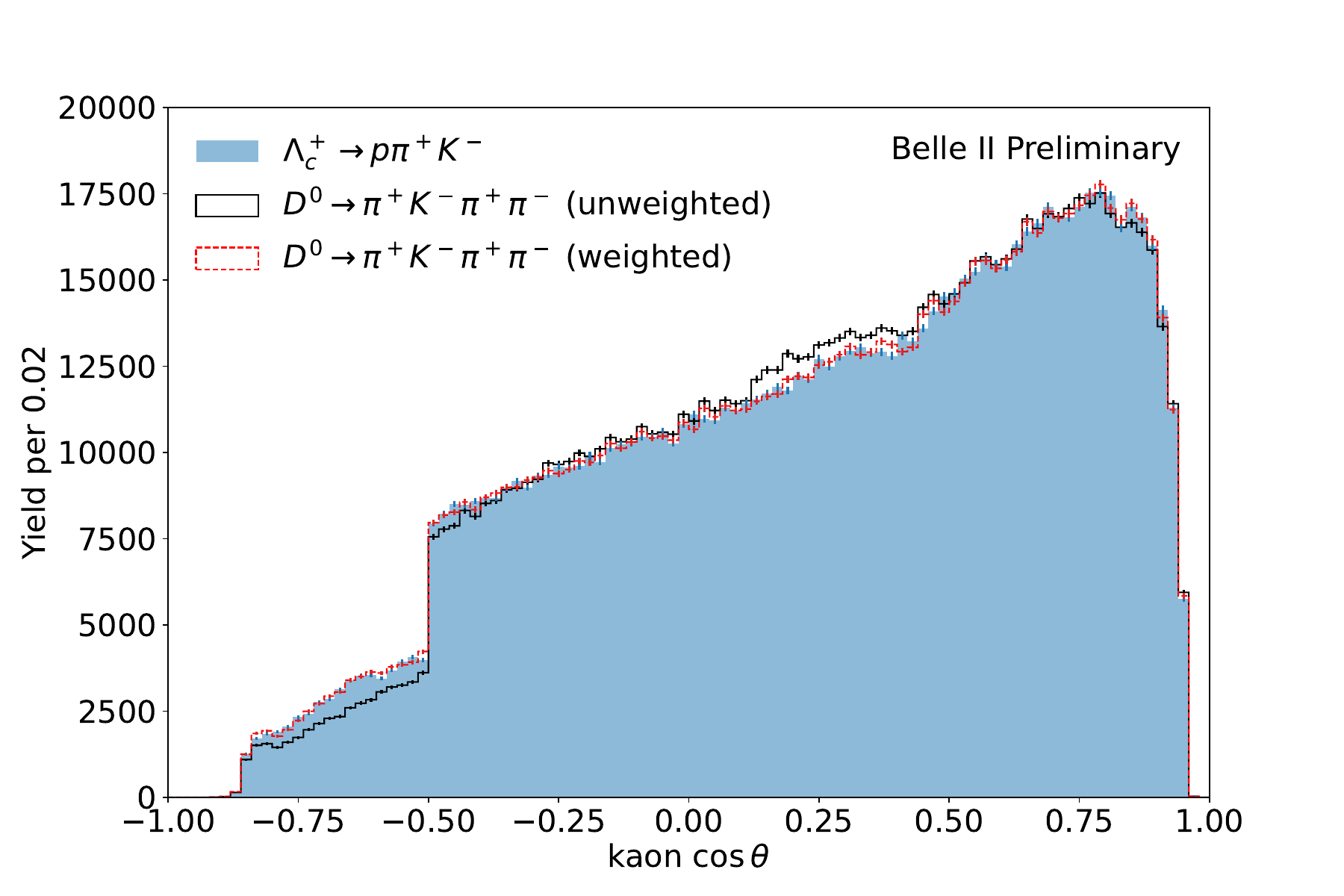}\\
\caption{Weighted and unweighted correct-candidate proton momentum~(left) and $\cos\theta$~(right) distributions for \LcTopKK and \LcTopKpi~(top) and \LcTopKpi and \DToKpipipi~(bottom) with selection criteria for \LcTopKK.}
\label{fig:LcToKK_weights}
\end{figure*}

\begin{figure*}[ht]
\centering
\includegraphics[width=0.45\linewidth]{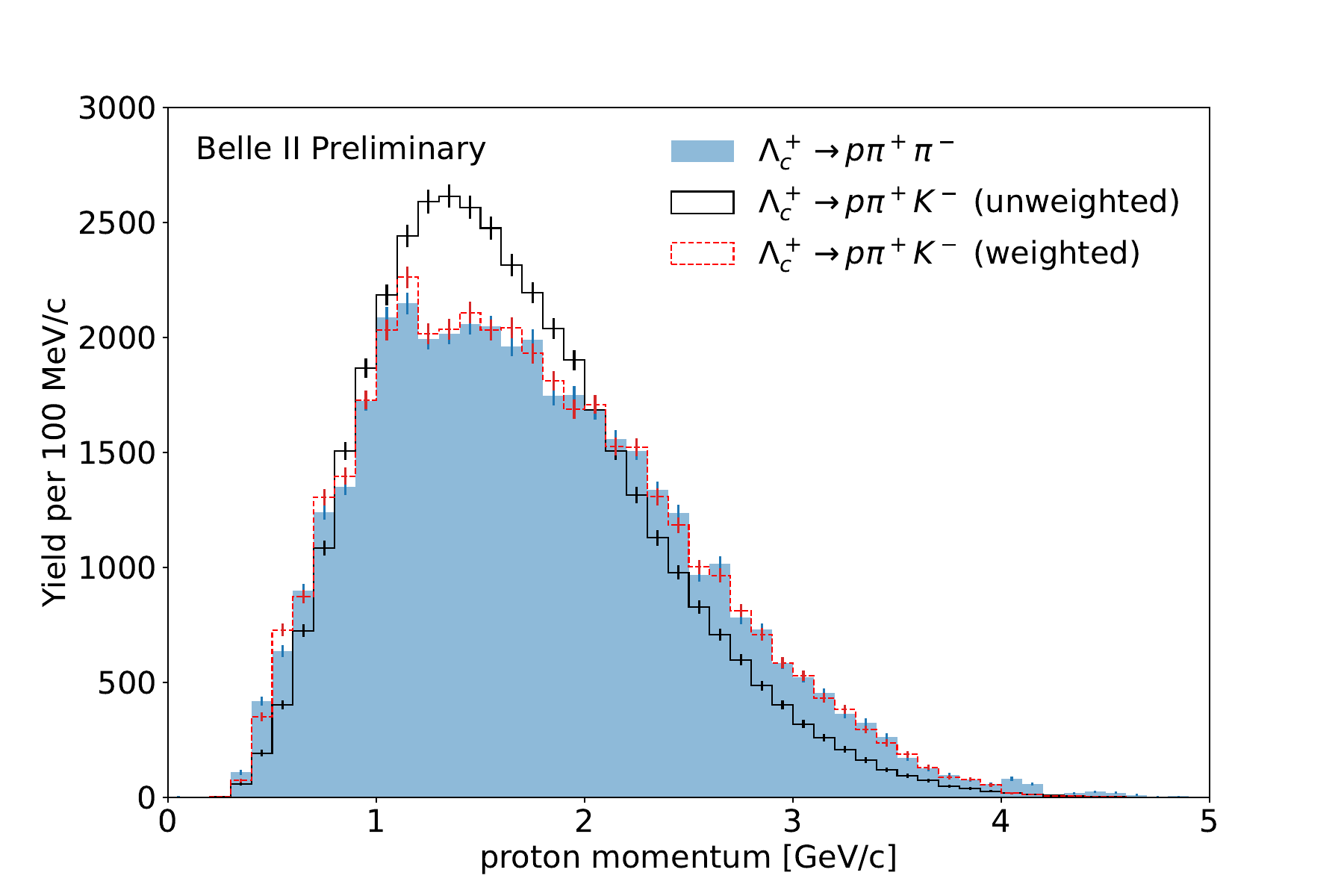}\hfil
\includegraphics[width=0.45\linewidth]{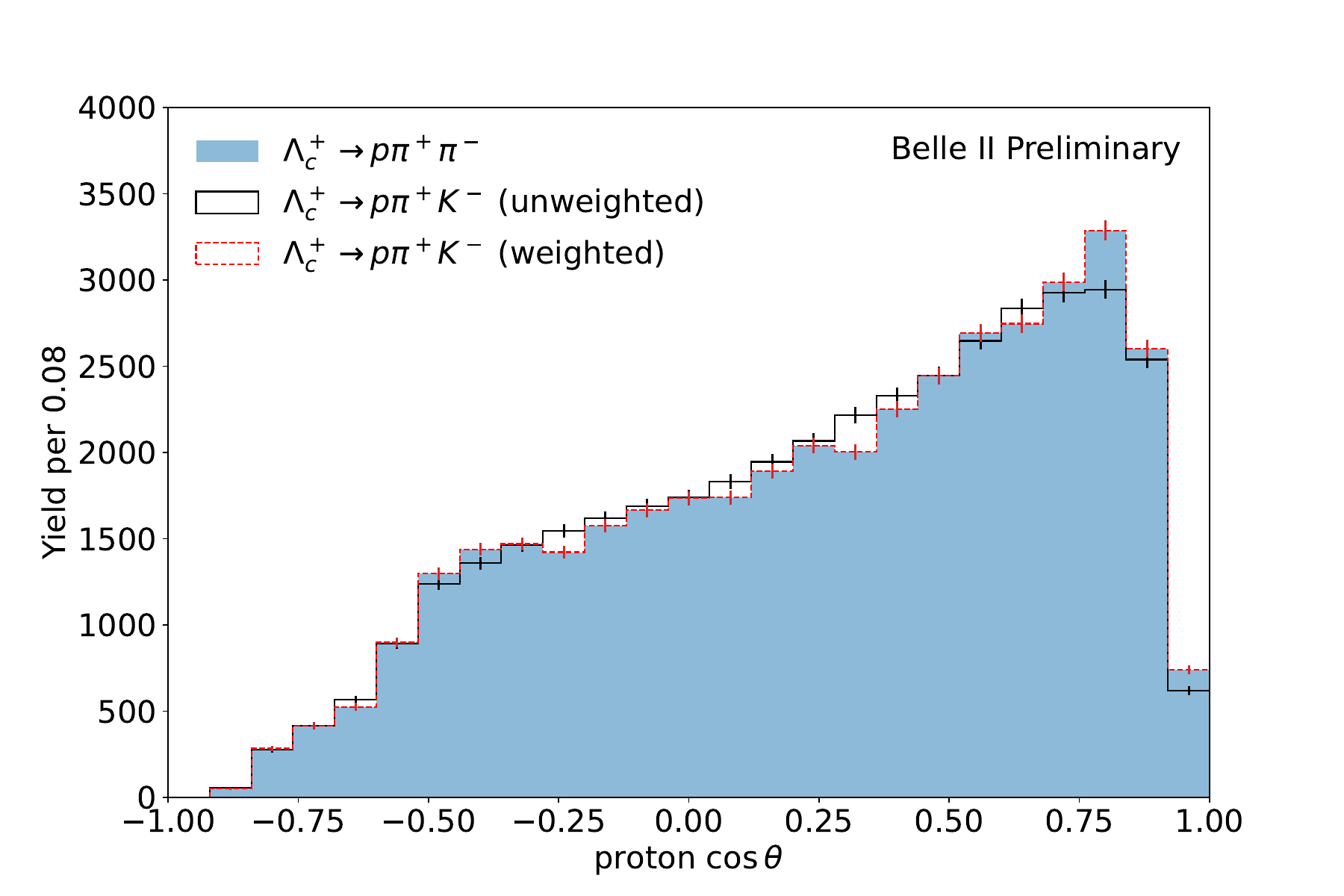}\\
\includegraphics[width=0.45\linewidth]{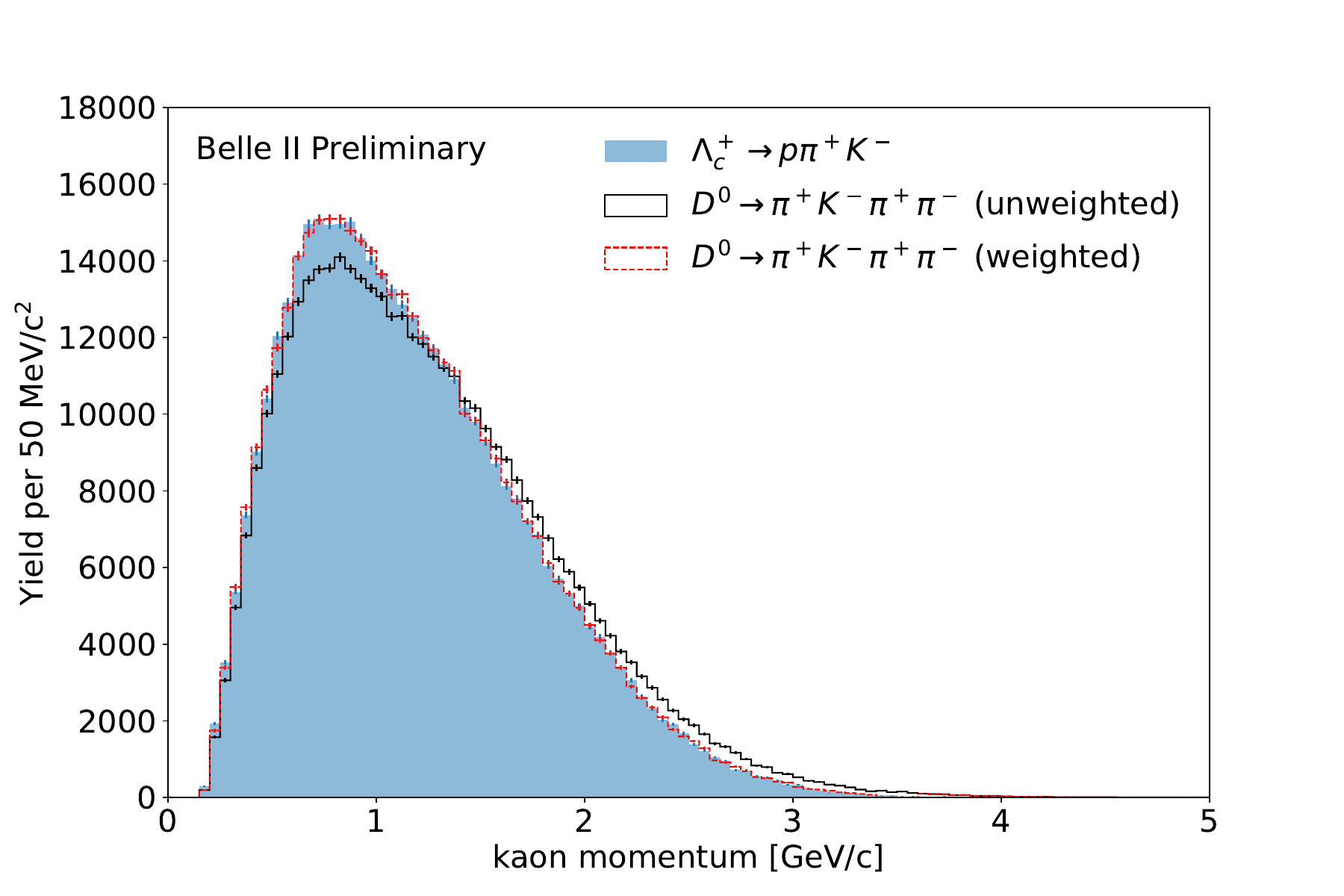}\hfil
\includegraphics[width=0.45\linewidth]{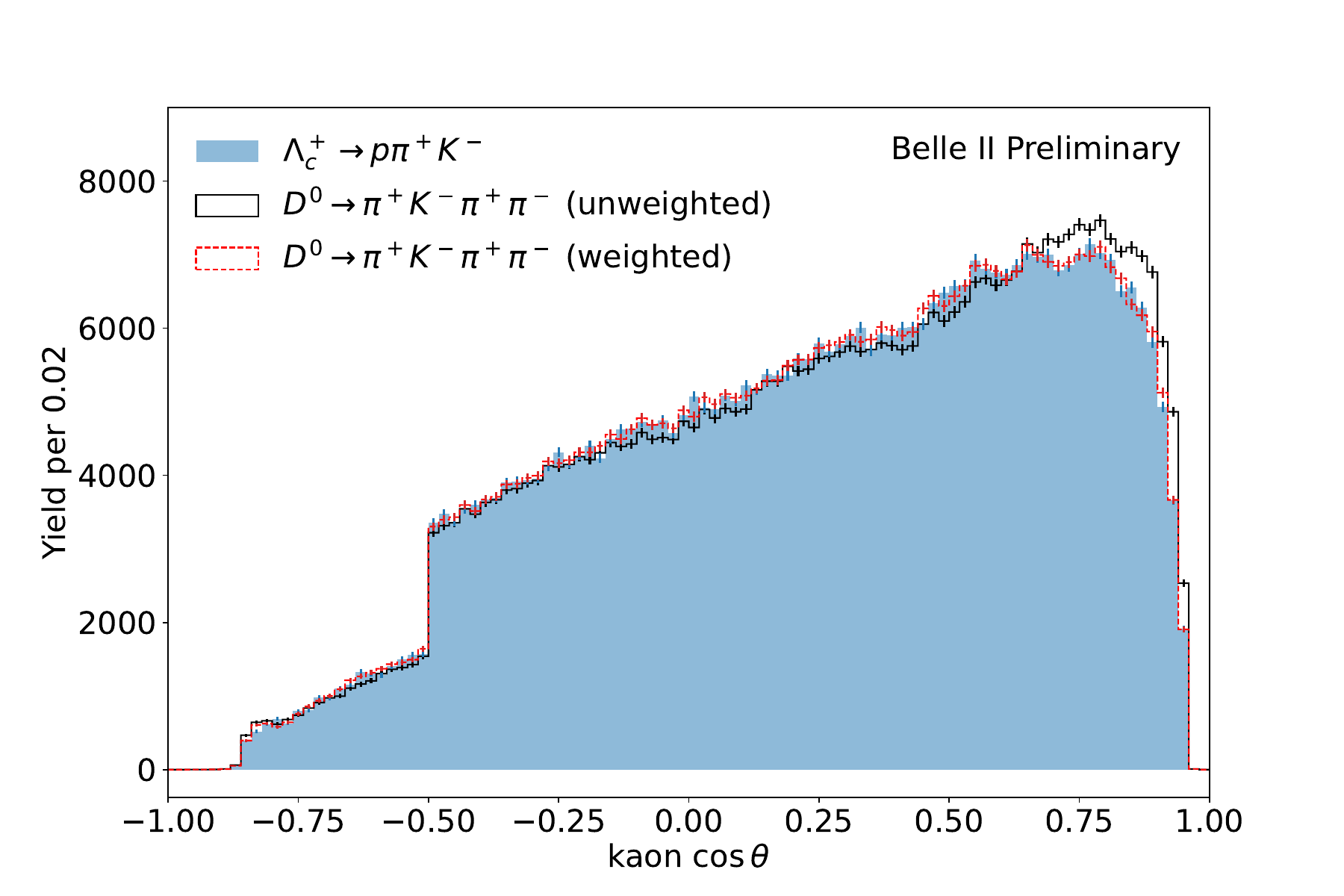}\\
\caption{Weighted and unweighted correct-candidate proton momentum~(left) and $\cos\theta$~(right) distributions for \LcToppipi and \LcTopKpi~(top) and \LcTopKpi and \DToKpipipi~(bottom) with selection criteria for \LcToppipi.}
\label{fig:LcTopipi_weights}
\end{figure*}

The kinematic weights for the \LcToShh and \LcTopKpi control channels are calculated from the momentum and $\cos\theta$ distributions of the final-state proton. The weights for the \DToKpipipi control channel are calculated from the momentum and $\cos\theta$ distributions of the kaon. Weighting in these distributions is sufficient to also equalize the \Sigp and \pip momentum and $\cos\theta$ distributions. We use the sPlot technique~\cite{Pivk:2004ty} to isolate the correct-candidate distributions, using the results of the yield fits. We first calculate weights to equalize the momentum distributions, apply those weights, and then calculate weights to equalize the $\cos\theta$ distributions. \Cref{fig:ScTohh_weights,fig:LcToKK_weights,fig:LcTopipi_weights} show these distributions with and without weighting. Small deviations remain in the momentum distributions, but the effect on the cancelation of detection asymmetries is negligible.

\begin{figure*}[ht]
\centering
\includegraphics[width=0.45\linewidth]{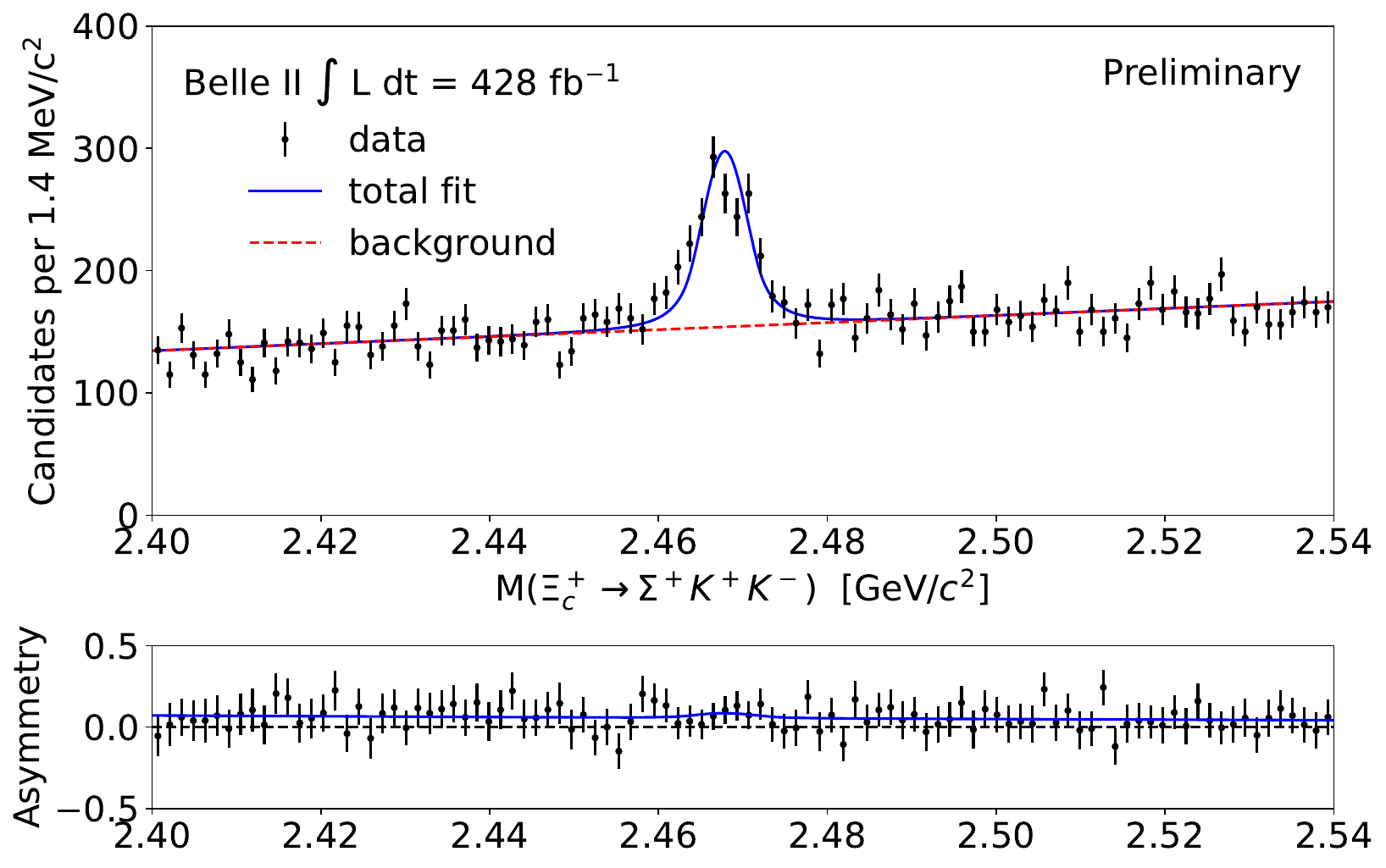}\hfil
\includegraphics[width=0.45\linewidth]{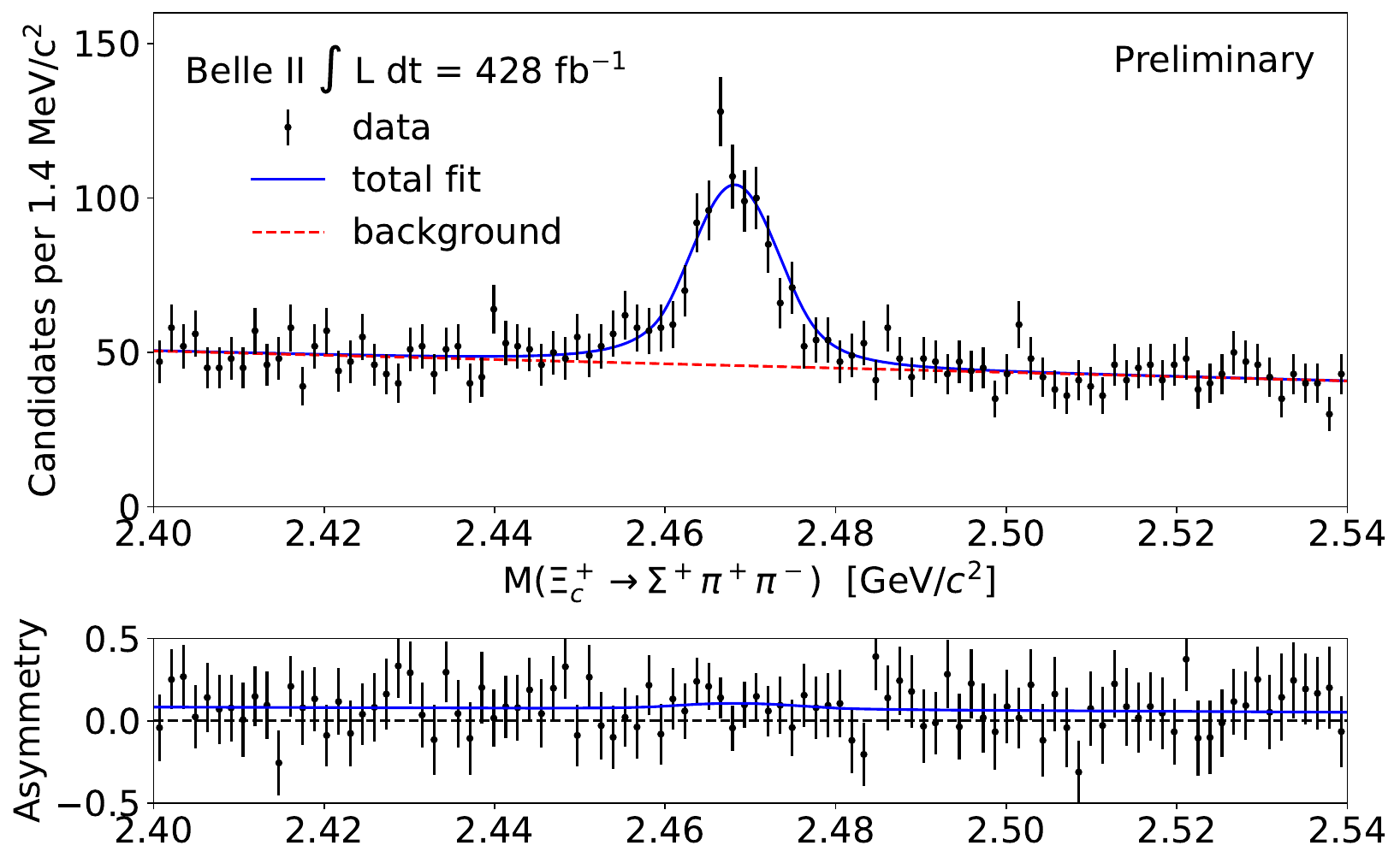}\\
\includegraphics[width=0.45\linewidth]{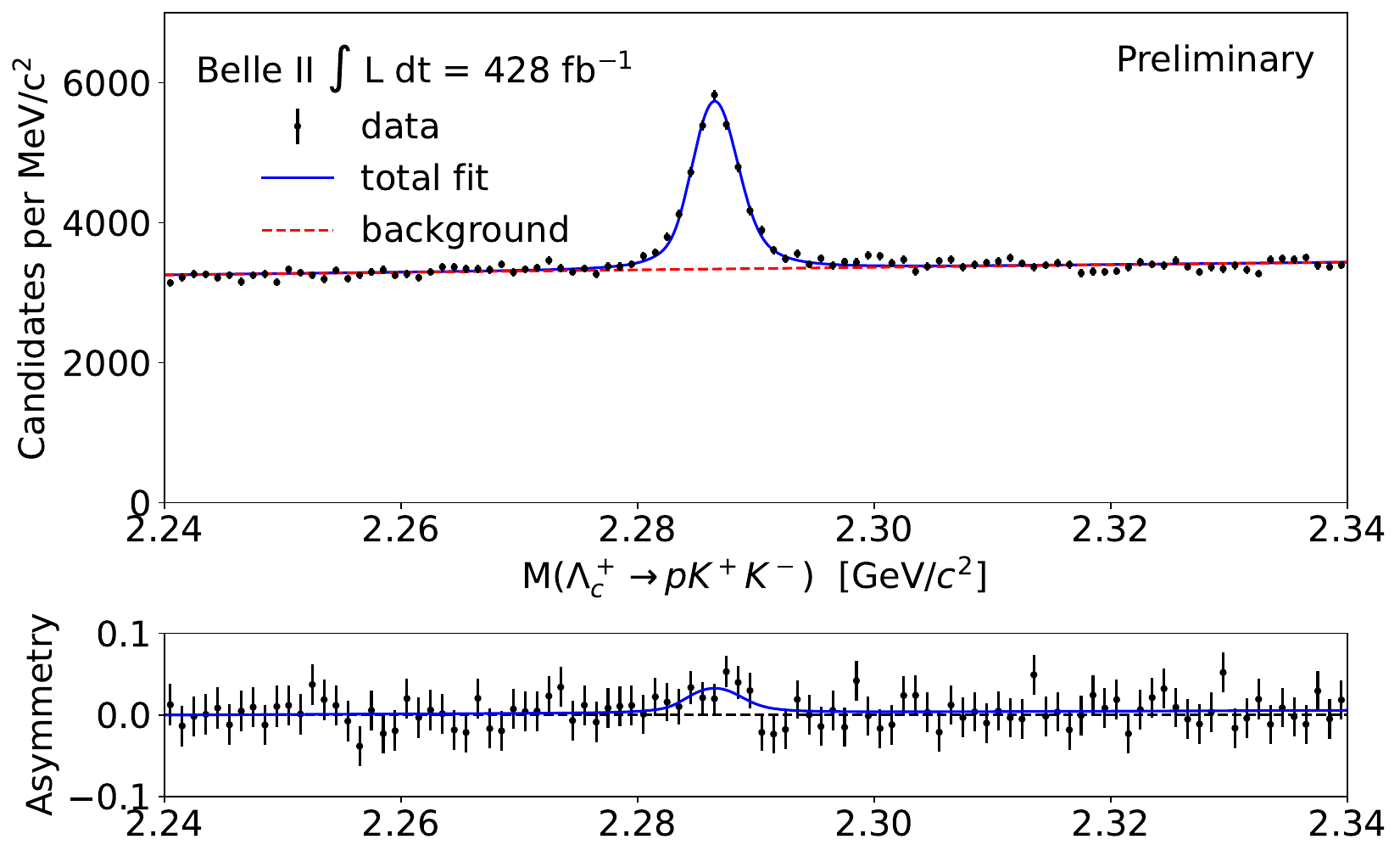}\hfil
\includegraphics[width=0.45\linewidth]{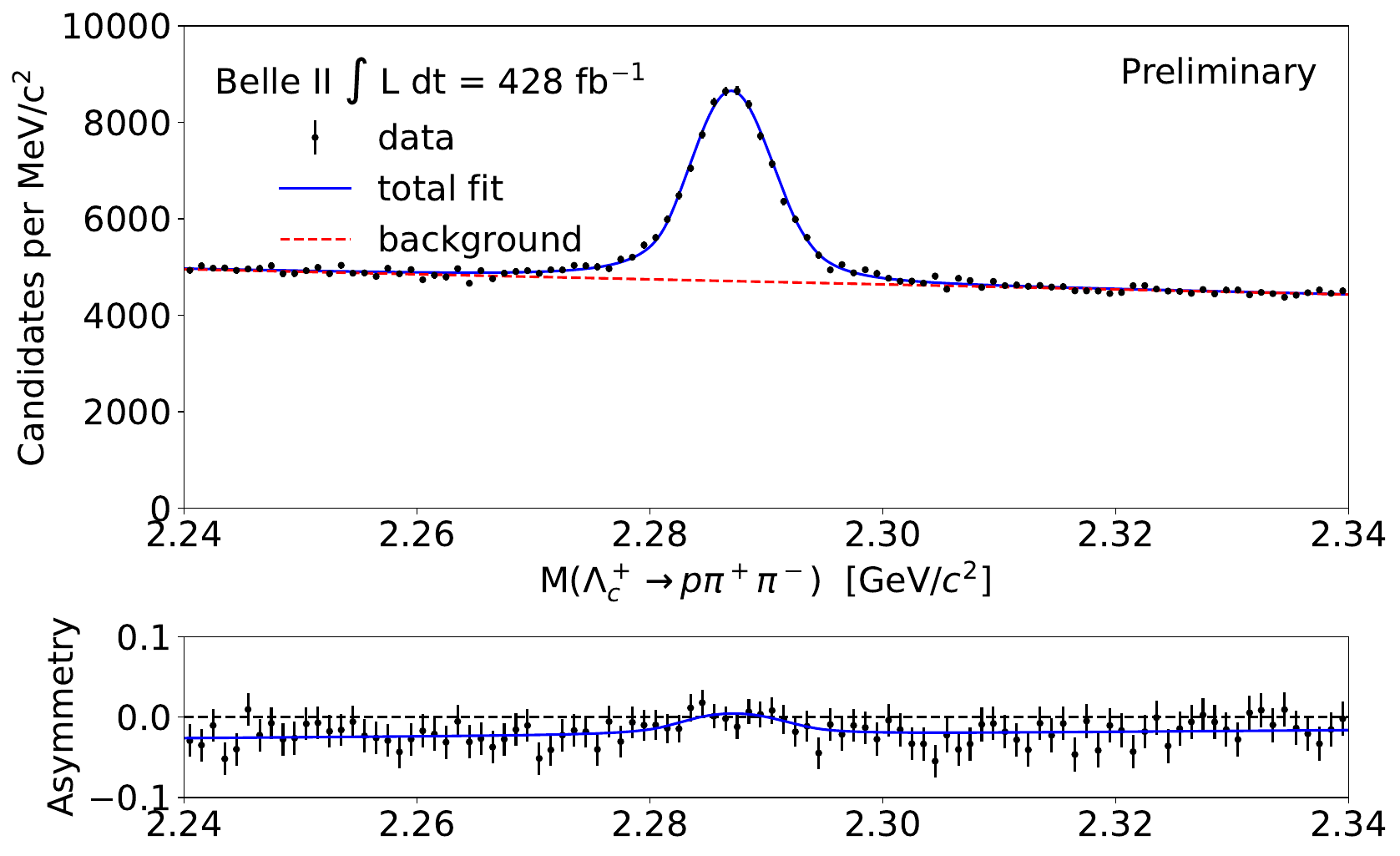}\\
\caption{Mass distributions for \XcToShh~(top) and \LcTophh~(bottom) candidates for $h=K$~(left) and $\pi$~(right) and the results of the fits, summing forward and backward contributions; and their averaged yield asymmetries as functions of mass, with fit projection overlaid.}
\label{fig:fit-signal}
\end{figure*}

\begin{figure*}[ht]
\centering
\includegraphics[width=0.45\linewidth]{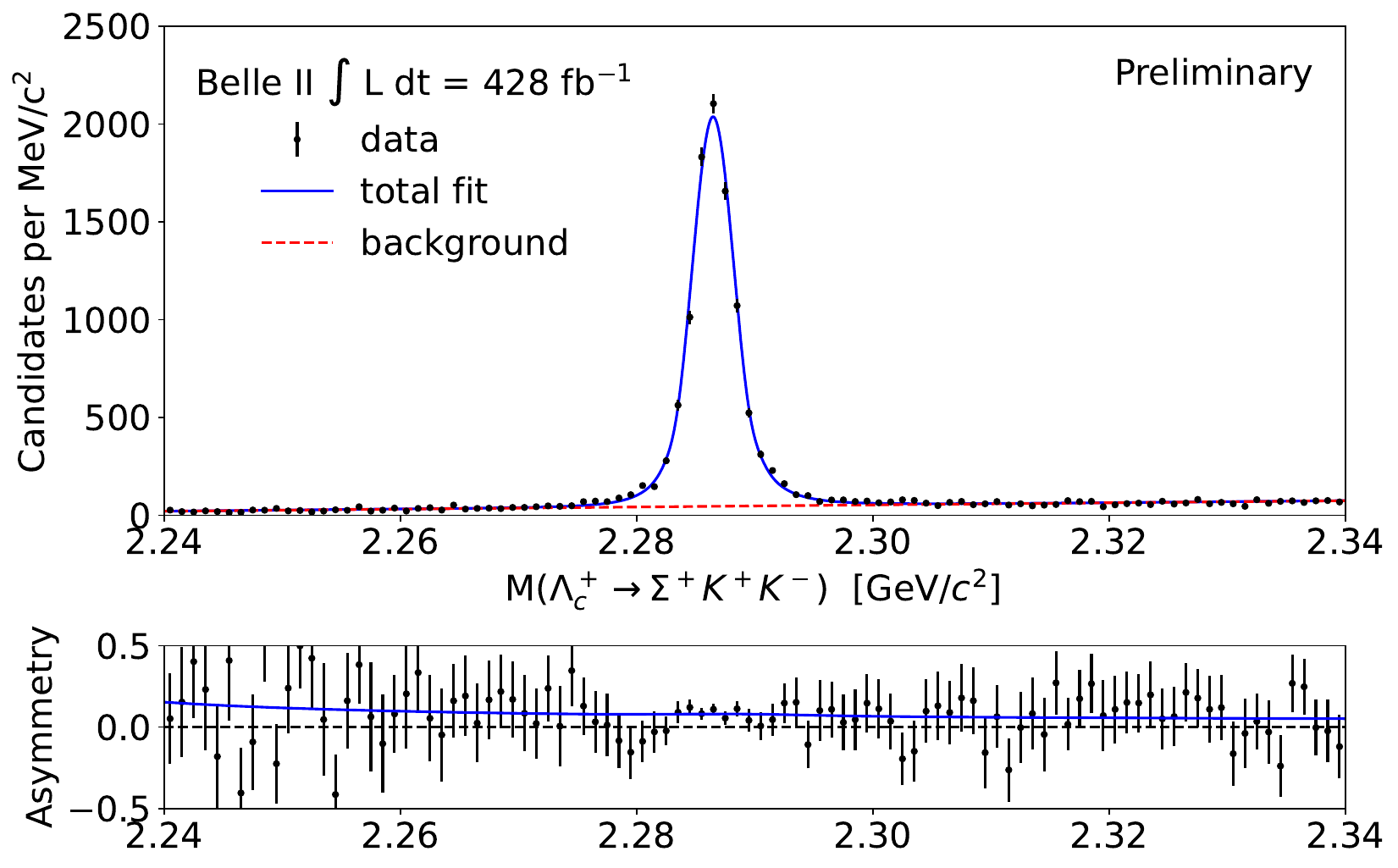}\hfil
\includegraphics[width=0.45\linewidth]{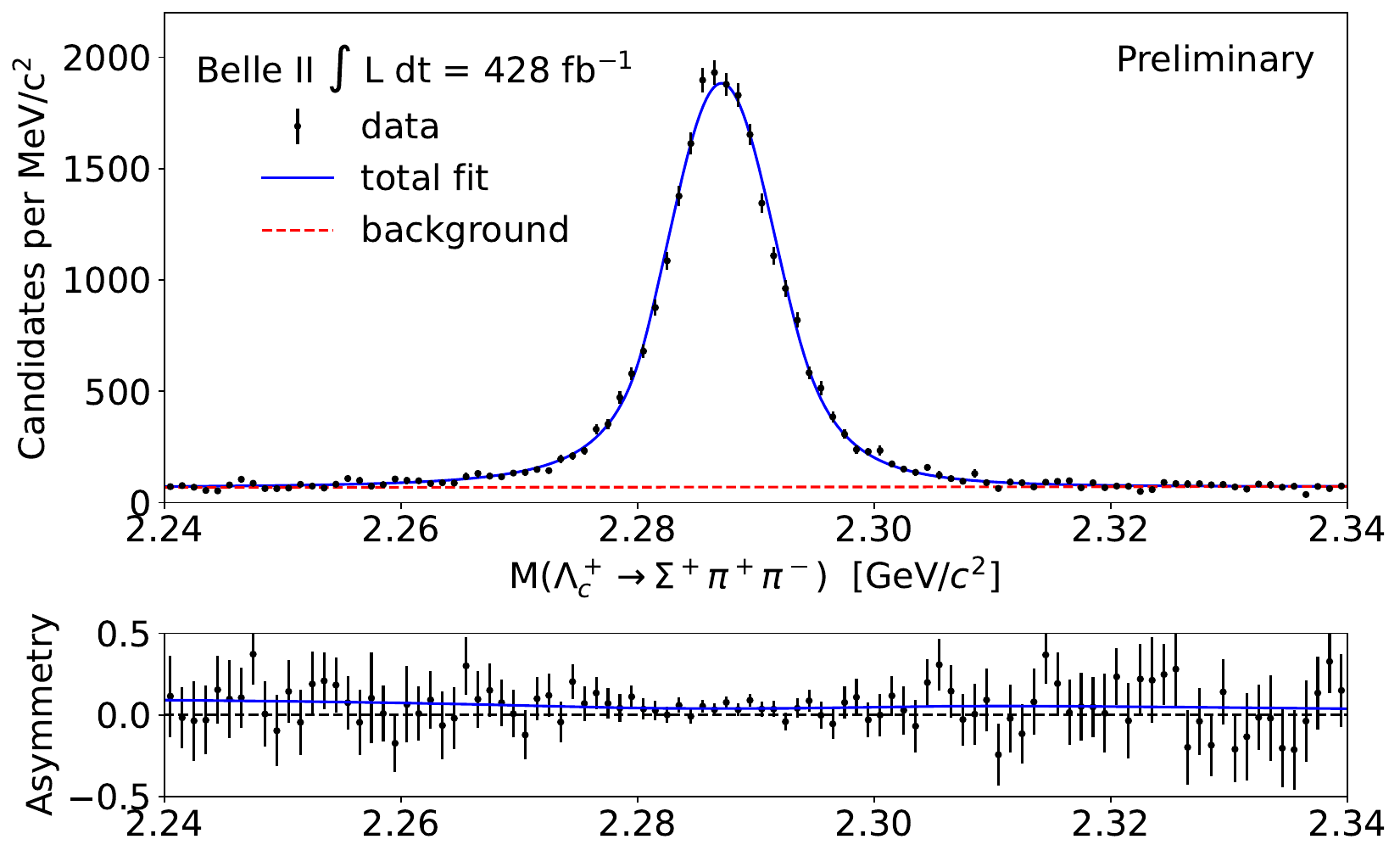}\\
\includegraphics[width=0.45\linewidth]{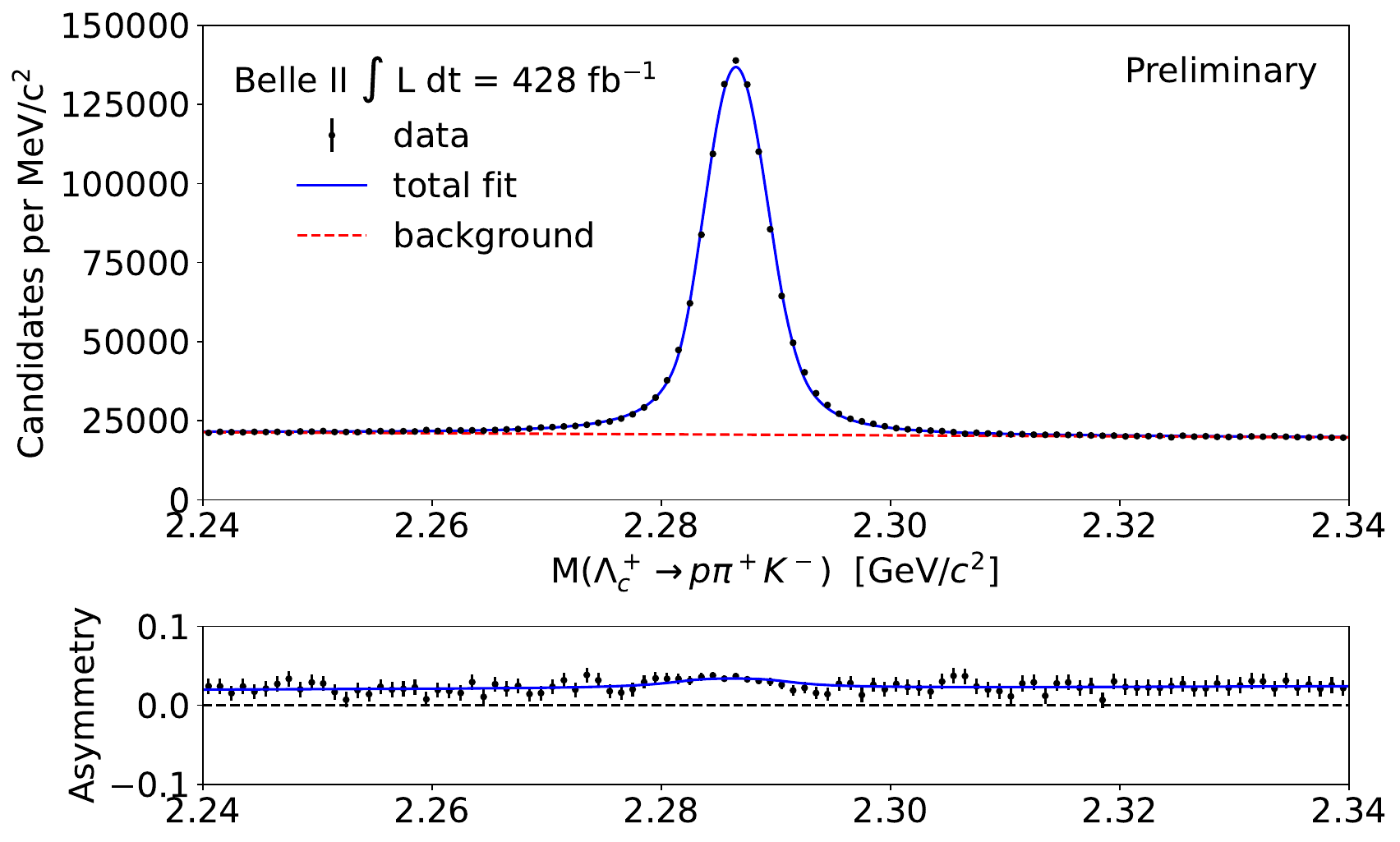}\hfil
\includegraphics[width=0.45\linewidth]{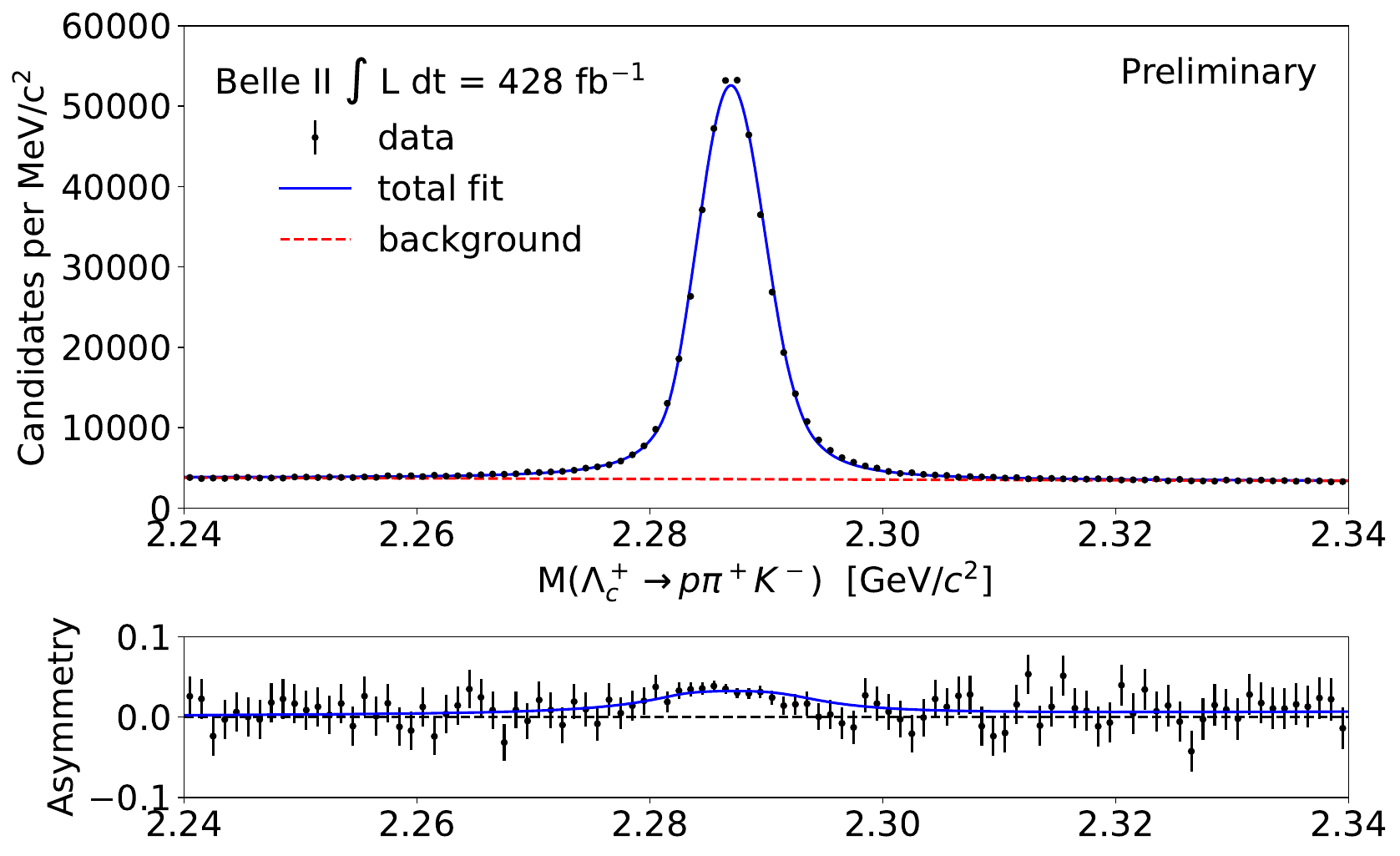}\\
\includegraphics[width=0.45\linewidth]{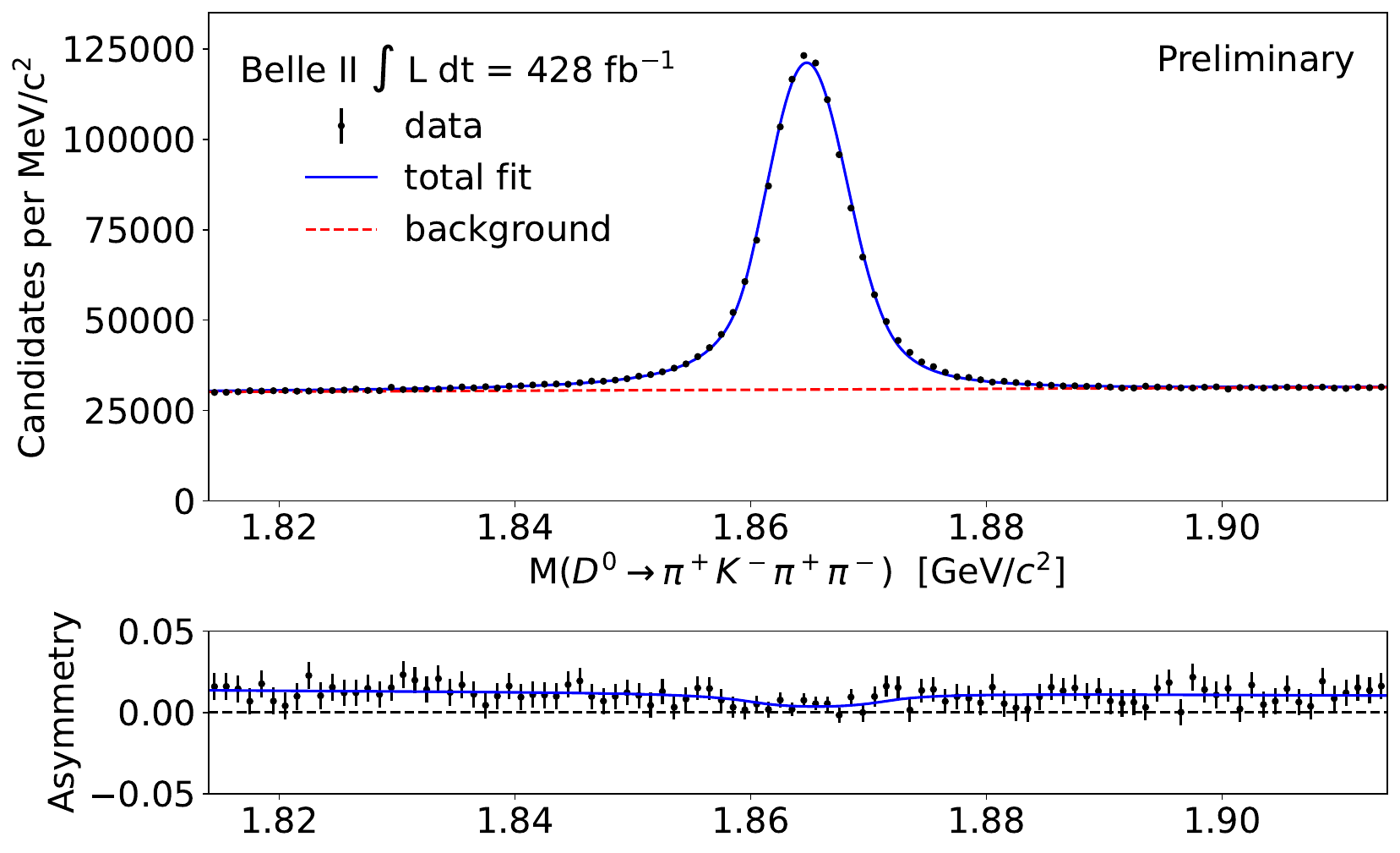}\hfil
\includegraphics[width=0.45\linewidth]{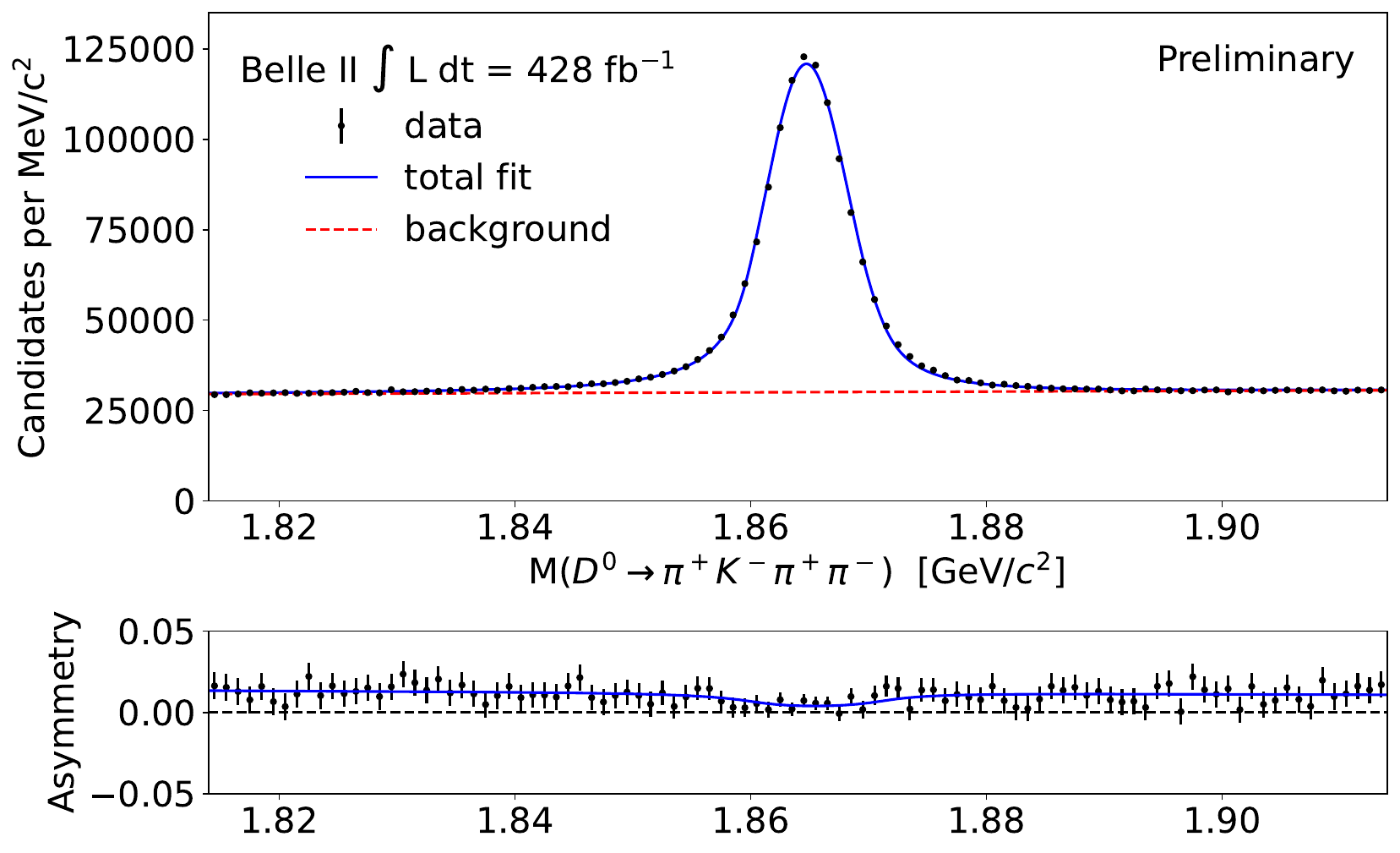}\\
\caption{Mass distributions for \LcToShh~(top), \LcTopKpi~(middle), and \DToKpipipi~(bottom) candidates with the selection criteria for \LcTopKK~(left) and \LcToppipi~(right) and the results of the fits, summing the forward and backward contributions; and their averaged yield asymmetries as functions of mass, with fit projection overlaid.}
\label{fig:fit-control}    
\end{figure*}

\begin{table*}[ht]
\centering
\caption{Yields (in $10^3$) and asymmetries (in \%) with statistical uncertainties. $\dagger$ and $\ddagger$ indicate candidates selected and kinematically weighted for the \LcTopKK and \LcToppipi modes, respectively.\label{tab:acp}}
\begin{tabular}{lrccc}
\toprule
Decay mode & Yield $\quad$ & Forward \Araw & Backward \Araw & $\Araw'$ \\
\midrule
\XcToSKK   & $0.78\pm0.05$ & $13.0\pm9.2\phantom{0}$ & $10.5\pm9.2\phantom{0}$ & $11.7\pm6.5\phantom{0}$ \\
\LcToSKK   & $9.9\pm0.1\phantom{0}$ & $10.9\pm1.5\phantom{0}$ & $5.3\pm1.6$ & $8.1\pm1.1$\\
\midrule
\XcToSpipi & $0.62\pm0.04$ & $17.0\pm10.0$ & $9.7\pm8.9$ & $13.3\pm6.8\phantom{0}$\\
\LcToSpipi & $23.4\pm0.2\phantom{0}$ & $7.4\pm1.0$ & $0.2\pm1.0$ & $3.8 \pm 0.8$\\
\midrule
\LcTopKK   & $13.6\pm0.2\phantom{0}$ & $9.3\pm2.2$ & $5.5\pm2.4$ & $7.4\pm1.7$\\
$^\dagger\LcTopKpi$ & $955.0\pm1.3\phantom{0}$ & $5.6\pm0.2$ & $1.6\pm0.2$ & $3.6\pm0.1$\\
$^\dagger\DToKpipipi$ & $928.0\pm1.4\phantom{0}$ & $1.6\pm0.2$ & $-1.5\pm0.2\phantom{-}$ & $0.1\pm0.2$\\
\midrule
\LcToppipi & $40.5\pm0.4\phantom{0}$ & $5.8\pm1.3$ & $1.5\pm1.4$ & $3.6 \pm 0.9$\\
$^\ddagger\LcTopKpi$ & $410.3\pm0.7\phantom{0}$ & $5.4\pm0.3$ & $1.5\pm0.3$ & $3.4\pm0.2$\\
$^\ddagger\DToKpipipi$ & $925.2\pm1.4\phantom{0}$ & $1.6\pm0.2$ & $-1.3\pm0.2\phantom{-}$ & $0.1\pm0.2$\\
\bottomrule
\end{tabular}
\end{table*}

We then repeat the fits with the weighting to determine the yields and the asymmetries of the control modes. \Cref{fig:fit-signal,fig:fit-control} show the mass distributions of the candidates for each decay channel, the results of the final fit, and the resulting averaged yield asymmetries as functions of the masses. \Cref{tab:acp} lists the measured yields and forward, backward, and averaged yield asymmetries. The measured asymmetries in data sidebands, which indicate the amount of asymmetry due to backgrounds, agree with the simulation.

We validate this procedure by fitting to simulated distributions generated by sampling from models fit to the real data, and with different input asymmetry values. We find no biases. 

\section{Systematic uncertainties}
\label{sec:systematics}

We repeat the fits to determine yields using a sum of Johnson's $S_U$ function~\cite{Johnson} and a Gaussian function in place of the double-sided Crystal Ball function for correct candidates and a second-order polynomial in place of the straight line for background (each separately). The results are consistent and we take the differences in quadrature of the asymmetry uncertainties as systematic uncertainties from the choice of the fit model.

We repeat the analysis with candidate weights using truth-matched candidates in simulation instead of using sPlot. We take the changes in results as systematic uncertainties arising from the weighting scheme. We also try determining the weights to equate the transverse-momentum distributions instead of the full-momentum distributions; the results change negligibly.

Since the detector acceptance depends on $\cos\theta$, a detection asymmetry due to the forward-backward asymmetry of charmed-baryon production may remain after averaging the forward and backward asymmetries. We estimate its effect by comparing the results of analyzing simulated data with and without candidates weighted to have identical $|\cos\theta|$ distributions for correct candidates (isolated using sPlot) in the forward and backward regions. The results are consistent. We take the differences as systematic uncertainties arising from potentially imperfect cancelation of $A\sub{p}$. We also compared the results of analyzing real data with and without weights, determined according to the same weight function as for simulated data, and found consistent values.

We assumed there is no asymmetry in detecting the \hphm pair from the \Xc and \Lc decays. To estimate the effects of such asymmetries, which may arise from intermediate states in the decays, we analyze data simulated to have exaggerated differences~(compared to the real data) in the $h^+$ and $h^-$ momentum distributions. The relative difference in average momentum of the two hadrons before and after modification ranges from 8\% to 17\%, depending on the channel. We take the differences in quadrature of the asymmetry uncertainties with and without this modification as systematic uncertainties due to a potential \hphm detection asymmetry, $A\sub{d}(\hphm)$.

We have also assumed there is no asymmetry in detecting the $\pip\pim$ pair in the \Dz decay. To estimate the effect of such an asymmetry, we compare the yield asymmetries in data for \DToKpipipi and \DToKpi, with the candidates of the four-body decay weighted so that the momentum and $\cos\theta$ distributions of its kaon and a randomly-selected \pip match those of the two-body decay. The weighted yield asymmetry for the four-body decay, $(-1.1\pm0.8)\%$, is consistent with that of the two-body decay, $(-0.4\pm0.2)\%$, showing that resonant structure in \DToKpipipi does not significantly bias the subtraction of the $\pi^+K^-$ detection asymmetry. We assign no systematic uncertainty for this potential effect.

\Cref{tab:systematics} lists these systematic uncertainties, their quadrature sum, and the statistical uncertainties for each signal decay mode.

\begin{table}[ht]
\centering
\caption{Uncertainties on \Acp (in \%).\label{tab:systematics}}
\begin{tabular}{l*{4}{c}}
\toprule
        Source
        & \multicolumn{2}{c}{\Xc}
        & \multicolumn{2}{c}{\Lc}
        \\

        \cmidrule(lr){2-3}
        \cmidrule(lr){4-5}
        
        & \SKK
        & \Spipi
        & \pKK
        & \ppipi
        \\

        \midrule

        Fit model
        & 0.4
        & 0.4
        & 0.4
        & 0.1
        \\
    
        Weighting
        & 0.1
        & 0.2
        & 0.3
        & 0.1
        \\

        Residual $A\sub{p}$
        & 0.2
        & 0.1
        & 0.1
        & 0.1
        \\
        
        $A\sub{d}(\hphm)$
        & 0.4
        & 0.2
        & 0.4
        & 0.1
        \\
        
        \midrule

        Total systematic
        & \AcpSKKSyst
        & \AcpSpipiSyst
        & \AcppKKSyst
        & \AcpppipiSyst
        \\
        
        Statistical
        & \AcpSKKStat
        & \AcpSpipiStat
        & \AcppKKStat
        & \AcpppipiStat
        \\

        \bottomrule
\end{tabular}
\end{table}

\section{Results and conclusions}

Using \lumi of \epem collisions collected by \belletwo, we measure
\begin{alignat}{1}
    \Acp(\XcToSKK)   &= (\AcpSKK   \pm \AcpSKKStat   \pm \AcpSKKSyst)\%,\\
    \Acp(\XcToSpipi) &= (\AcpSpipi \pm \AcpSpipiStat \pm \AcpSpipiSyst)\%,\\
    \Acp(\LcTopKK)   &= (\AcppKK   \pm \AcppKKStat   \pm \AcppKKSyst)\%,\\
    \Acp(\LcToppipi) &= (\Acpppipi \pm \AcpppipiStat \pm \AcpppipiSyst)\%,   
\end{alignat}
where the first uncertainties are statistical and the second ones systematic. These results agree with \CP symmetry. Their $U$-spin sums are
\ifthenelse{\boolean{twocolumnstyle}}{
    \begin{multline}
        \Acp(\XcToSpipi) + \Acp(\LcTopKK) \\ = (13.4 \pm 7.0\pm 0.9)\%,
    \end{multline}
    \begin{multline}
        \Acp(\XcToSKK) + \Acp(\LcToppipi) \\ = (\phantom{0}4.0 \pm 6.6\pm 0.7)\%,
    \end{multline}
}{
    \begin{alignat}{3}
        &\Acp(\XcToSpipi) &&\, + \, \Acp(\LcTopKK)   &&= (13.4 \pm 7.0\pm 0.9)\%,\\
        &\Acp(\XcToSKK)   &&\, + \, \Acp(\LcToppipi) &&= (\phantom{0}4.0 \pm 6.6\pm 0.7)\%,
    \end{alignat}
 }
consistent with $U$-spin symmetry. These are the world's first \Acp measurements for individual hadronic three-body charmed-baryon decays. Their uncertainties are predominantly statistical in origin, so future measurements using more data collected by \belletwo will be important for precisely searching for \CP violation and testing $U$-spin sum rules.

\section*{Acknowledgements}
This work, based on data collected using the Belle II detector, which was built and commissioned prior to March 2019,
was supported by
Higher Education and Science Committee of the Republic of Armenia Grant No.~23LCG-1C011;
Australian Research Council and Research Grants
No.~DP200101792, 
No.~DP210101900, 
No.~DP210102831, 
No.~DE220100462, 
No.~LE210100098, 
and
No.~LE230100085; 
Austrian Federal Ministry of Education, Science and Research,
Austrian Science Fund (FWF) Grants
DOI:~10.55776/P34529,
DOI:~10.55776/J4731,
DOI:~10.55776/J4625,
DOI:~10.55776/M3153,
and
DOI:~10.55776/PAT1836324,
and
Horizon 2020 ERC Starting Grant No.~947006 ``InterLeptons'';
Natural Sciences and Engineering Research Council of Canada, Digital Research Alliance of Canada, and Canada Foundation for Innovation;
National Key R\&D Program of China under Contract No.~2024YFA1610503,
and
No.~2024YFA1610504
National Natural Science Foundation of China and Research Grants
No.~11575017,
No.~11761141009,
No.~11705209,
No.~11975076,
No.~12135005,
No.~12150004,
No.~12161141008,
No.~12405099,
No.~12475093,
and
No.~12175041,
and Shandong Provincial Natural Science Foundation Project~ZR2022JQ02;
the Czech Science Foundation Grant No. 22-18469S,  Regional funds of EU/MEYS: OPJAK
FORTE CZ.02.01.01/00/22\_008/0004632 
and
Charles University Grant Agency project No. 246122;
European Research Council, Seventh Framework PIEF-GA-2013-622527,
Horizon 2020 ERC-Advanced Grants No.~267104 and No.~884719,
Horizon 2020 ERC-Consolidator Grant No.~819127,
Horizon 2020 Marie Sklodowska-Curie Grant Agreement No.~700525 ``NIOBE''
and
No.~101026516,
and
Horizon 2020 Marie Sklodowska-Curie RISE project JENNIFER2 Grant Agreement No.~822070 (European grants);
L'Institut National de Physique Nucl\'{e}aire et de Physique des Particules (IN2P3) du CNRS
and
L'Agence Nationale de la Recherche (ANR) under Grant No.~ANR-23-CE31-0018 (France);
BMFTR, DFG, HGF, MPG, and AvH Foundation (Germany);
Department of Atomic Energy under Project Identification No.~RTI 4002,
Department of Science and Technology,
and
UPES SEED funding programs
No.~UPES/R\&D-SEED-INFRA/17052023/01 and
No.~UPES/R\&D-SOE/20062022/06 (India);
Israel Science Foundation Grant No.~2476/17,
U.S.-Israel Binational Science Foundation Grant No.~2016113, and
Israel Ministry of Science Grant No.~3-16543;
Istituto Nazionale di Fisica Nucleare and the Research Grants BELLE2,
and
the ICSC – Centro Nazionale di Ricerca in High Performance Computing, Big Data and Quantum Computing, funded by European Union – NextGenerationEU;
Japan Society for the Promotion of Science, Grant-in-Aid for Scientific Research Grants
No.~16H03968,
No.~16H03993,
No.~16H06492,
No.~16K05323,
No.~17H01133,
No.~17H05405,
No.~18K03621,
No.~18H03710,
No.~18H05226,
No.~19H00682, 
No.~20H05850,
No.~20H05858,
No.~22H00144,
No.~22K14056,
No.~22K21347,
No.~23H05433,
No.~26220706,
and
No.~26400255,
and
the Ministry of Education, Culture, Sports, Science, and Technology (MEXT) of Japan;  
National Research Foundation (NRF) of Korea Grants
No.~2021R1-F1A-1064008, 
No.~2022R1-A2C-1003993,
No.~2022R1-A2C-1092335,
No.~RS-2016-NR017151,
No.~RS-2018-NR031074,
No.~RS-2021-NR060129,
No.~RS-2023-00208693,
No.~RS-2024-00354342
and
No.~RS-2025-02219521,
Radiation Science Research Institute,
Foreign Large-Size Research Facility Application Supporting project,
the Global Science Experimental Data Hub Center, the Korea Institute of Science and
Technology Information (K25L2M2C3 ) 
and
KREONET/GLORIAD;
Universiti Malaya RU grant, Akademi Sains Malaysia, and Ministry of Education Malaysia;
Frontiers of Science Program Contracts
No.~FOINS-296,
No.~CB-221329,
No.~CB-236394,
No.~CB-254409,
and
No.~CB-180023, and SEP-CINVESTAV Research Grant No.~237 (Mexico);
the Polish Ministry of Science and Higher Education and the National Science Center;
the Ministry of Science and Higher Education of the Russian Federation
and
the HSE University Basic Research Program, Moscow;
University of Tabuk Research Grants
No.~S-0256-1438 and No.~S-0280-1439 (Saudi Arabia), and
Researchers Supporting Project number (RSPD2025R873), King Saud University, Riyadh,
Saudi Arabia;
Slovenian Research Agency and Research Grants
No.~J1-50010
and
No.~P1-0135;
Ikerbasque, Basque Foundation for Science,
State Agency for Research of the Spanish Ministry of Science and Innovation through Grant No. PID2022-136510NB-C33, Spain,
Agencia Estatal de Investigacion, Spain
Grant No.~RYC2020-029875-I
and
Generalitat Valenciana, Spain
Grant No.~CIDEGENT/2018/020;
The Knut and Alice Wallenberg Foundation (Sweden), Contracts No.~2021.0174, No.~2021.0299, and No.~2023.0315;
National Science and Technology Council,
and
Ministry of Education (Taiwan);
Thailand Center of Excellence in Physics;
TUBITAK ULAKBIM (Turkey);
National Research Foundation of Ukraine, Project No.~2020.02/0257,
and
Ministry of Education and Science of Ukraine;
the U.S. National Science Foundation and Research Grants
No.~PHY-1913789 
and
No.~PHY-2111604, 
and the U.S. Department of Energy and Research Awards
No.~DE-AC06-76RLO1830, 
No.~DE-SC0007983, 
No.~DE-SC0009824, 
No.~DE-SC0009973, 
No.~DE-SC0010007, 
No.~DE-SC0010073, 
No.~DE-SC0010118, 
No.~DE-SC0010504, 
No.~DE-SC0011784, 
No.~DE-SC0012704, 
No.~DE-SC0019230, 
No.~DE-SC0021274, 
No.~DE-SC0021616, 
No.~DE-SC0022350, 
No.~DE-SC0023470; 
and
the Vietnam Academy of Science and Technology (VAST) under Grants
No.~NVCC.05.02/25-25
and
No.~DL0000.05/26-27.

These acknowledgements are not to be interpreted as an endorsement of any statement made
by any of our institutes, funding agencies, governments, or their representatives.

We thank the SuperKEKB team for delivering high-luminosity collisions;
the KEK cryogenics group for the efficient operation of the detector solenoid magnet and IBBelle on site;
the KEK Computer Research Center for on-site computing support; the NII for SINET6 network support;
and the raw-data centers hosted by BNL, DESY, GridKa, IN2P3, INFN, 
and the University of Victoria.


\bibliographystyle{belle2}
\bibliography{references}

\end{document}